%% file: paper.tex
\documentclass[submission, Phys]{SciPost}
\pdfoutput=1
\usepackage{amsmath,amssymb}
\usepackage{booktabs,multirow,graphicx,tabularx,mathtools,slashed,url,hyperref}
\usepackage{doi}
\usepackage[maxfloats=100]{morefloats}
\usepackage{color,xcolor,braket,comment}
\usepackage[normalem]{ulem}
\usepackage{enumitem}
\usepackage{feynmp}

\makeatletter
\@ifundefined{pdfoutput}{}{\DeclareGraphicsRule{*}{mps}{*}{}}
\makeatother

\input{definitions}
\begin{document}
\begin{fmffile}{feynman}

\begin{center}{\Large \textbf{
Tagging Jets in Invisible Higgs Searches
}}\end{center}

\begin{center}
Anke Biek\"otter,
Fabian Keilbach,
Rhea Moutafis,
Tilman Plehn, 
Jennifer M. Thompson
\end{center}

\begin{center}
Institut f\"ur Theoretische Physik, Universit\"at Heidelberg, Germany
\end{center}


\section*{Abstract}
{\bf Searches for invisible Higgs decays in weak boson fusion are a
  well-known laboratory for jets and QCD studies. We present a series
  of results on tagging jets and central jet activity. First,
  precision analyses of the central jet activity require full control
  of single top production in some analyses. Second, the rate dependence on the size of
  the tagging jets is not limited to weak boson fusion. For the first
  time, we show how subjet information on the tagging jets and on the
  additional jet activity can be used to extract the Higgs signal. The
  additional observables relieve some of the pressure on other,
  critical observables.  Finally, we compare the performance of weak
  boson fusion and associated Higgs production.  }



\newpage
\vspace{10pt}
\noindent\rule{\textwidth}{1pt}
\tableofcontents\thispagestyle{fancy}
\noindent\rule{\textwidth}{1pt}
\vspace{10pt}

\section{Introduction}
\label{sec:intro}

The discovery of a fundamental Higgs boson~\cite{higgs,discovery}
guarantees the LHC a place among the great experimental efforts in
science and establishes perturbative gauge theory as the framework
that describes fundamental interactions in nature. Since then we have
been experiencing an essential change in research performed at
colliders. Quietly and against all odds, LHC physics has become a
field of experimental and theoretical precision physics. This is
closely related to a new way we think about QCD and jets. From being
regarded as a nuisance in going from lepton colliders to hadron
colliders, jets have become a powerful analysis object far beyond
simply describing partons from hard processes. Key directions in jet
physics include QCD radiations as well as jet sub-structure including
advanced machine learning methods. This defines a turning point, where
it is not obvious if by the end of the current run ATLAS and CMS
should still use jets as contained objects rather than an optional
link between jet-based observables and sub-structure
observables.\bigskip

In view of both its physics potential and these technical developments, one
of the most interesting LHC analyses is the search for Higgs decays
into invisible particles~\cite{inv_decay}. On the physics side, dark
matter is the big open question in particle physics cosmology. In the
Standard Model, Higgs decays to neutrinos are extremely rare. In
models for physics beyond the Standard Model, invisible Higgs decays
are a generic signature. One common structural element is the
super-renormalizable Higgs mass term, which allows any singlet field
to mix with the Higgs and which opens a renormalizable portal to a
hidden sector~\cite{portal}. This Higgs portal opens a wealth of
options for model building ranging from simple dark matter models to
more complicated unified
models~\cite{nmssm_inv,hpair_inv,non_thermal}. Whatever guides the
exact composition of such a hidden sector, a Higgs portal would always
show itself through an invisible decay width.\bigskip

On the phenomenological and analysis side, there are various
strategies to detect invisible Higgs decays at the LHC.  The classic
search strategy for invisible Higgs decays is based on
weak-boson-fusion (WBF) Higgs
production~\cite{eboli_zeppenfeld,spying,wbf_inv}. Two forward tagging
jets~\cite{taggingjets,fwm} can help with triggering, and the massive
$W$-propagators automatically provide a minimum amount of missing
energy from the Higgs decay. A general challenge in weak boson fusion
processes are the large QCD backgrounds, which can be controlled
through a comprehensive analysis of the tagging jet kinematics and the
suppressed hadronic activity in the central detector. It is worth
noting that the difference in jet structure of the WBF signal and the
QCD backgrounds follows from first-principles perturbative QCD.

Formally of the same perturbative order as the WBF production process,
but with a distinctive resonance structure, boosted Higgs production
in association with either a $W$ or a $Z$ boson will significantly add to the
LHC reach~\cite{vh_inv,vh_inv2}. A leptonic $Z$-decay not only
guarantees triggering, but also provides a powerful handle in reducing
all QCD backgrounds. On the other hand, the production rate in this
channel is has a significantly smaller rate than WBF Higgs production.
Finally, searches for invisible Higgs decays in $t\bar{t}H$
production~\cite{tth_inv} will be a challenge even at the
high-luminosity LHC, both statistically and systematically.

Experimental searches for invisible Higgs decays by CMS rely on weak
boson fusion~\cite{CMS_wbf_inv13}, $ZH$
production~\cite{CMS_zh_inv13}, and a combination of both production
processes~\cite{cms_comb,cms_comb_7813}. Similarly, in ATLAS there are
searches in weak boson fusion~\cite{atlas_wbf_inv8}, $ZH$
production~\cite{atlas_zh}, and a combination including the hadronic
$ZH$ channel~\cite{atlas8_inv_comb}.  The reach in terms of an
invisible branching ratio of a Standard-Model-like Higgs boson ranges
around 23\%, but is expected to reach the 2~...~3\% level at the end
of the high-luminosity run~\cite{spying}. To avoid unnecessary
assumptions about the Higgs production rate, an invisible Higgs search
is best included in a global Higgs and gauge sector
analysis~\cite{legacy}.  An additional improvement to the sub-percent
level can be reached in the same signatures at a future 100~TeV hadron
collider~\cite{nimatron}. Even more optimistic predictions for the
100~TeV hadron collider are typically based on ignoring the known
leading sources of uncertainties.\bigskip

In this study we focus on the WBF production channel and possible
improvements through an improved understanding of the tagging jets and
the central jet activity.  In Sec.~\ref{sec:wbf_irred} we start by
discussing some background issues which occur when we link the
irreducible $Z$+jets backgrounds and the $W$+jets backgrounds, in
particular the impact of single top production as a future key
background. Next, we focus on the structure of the tagging jets,
including the effects of a change in jet size
(Sec.~\ref{sec:wbf_jetsize}) and its quark vs gluon content
(Sec.~\ref{sec:wbf_jetcont}). The final, crucial question for this
signature in the high-luminosity LHC environment is how much it gets
degraded by stronger trigger requirements. In
Sec.~\ref{sec:wbf_trigger} we compare different triggering scenarios
with a $ZH$ benchmark analysis.

\section{V+jets backgrounds}
\label{sec:wbf_irred}

\begin{figure}[t]
\includegraphics[width=0.32\textwidth]{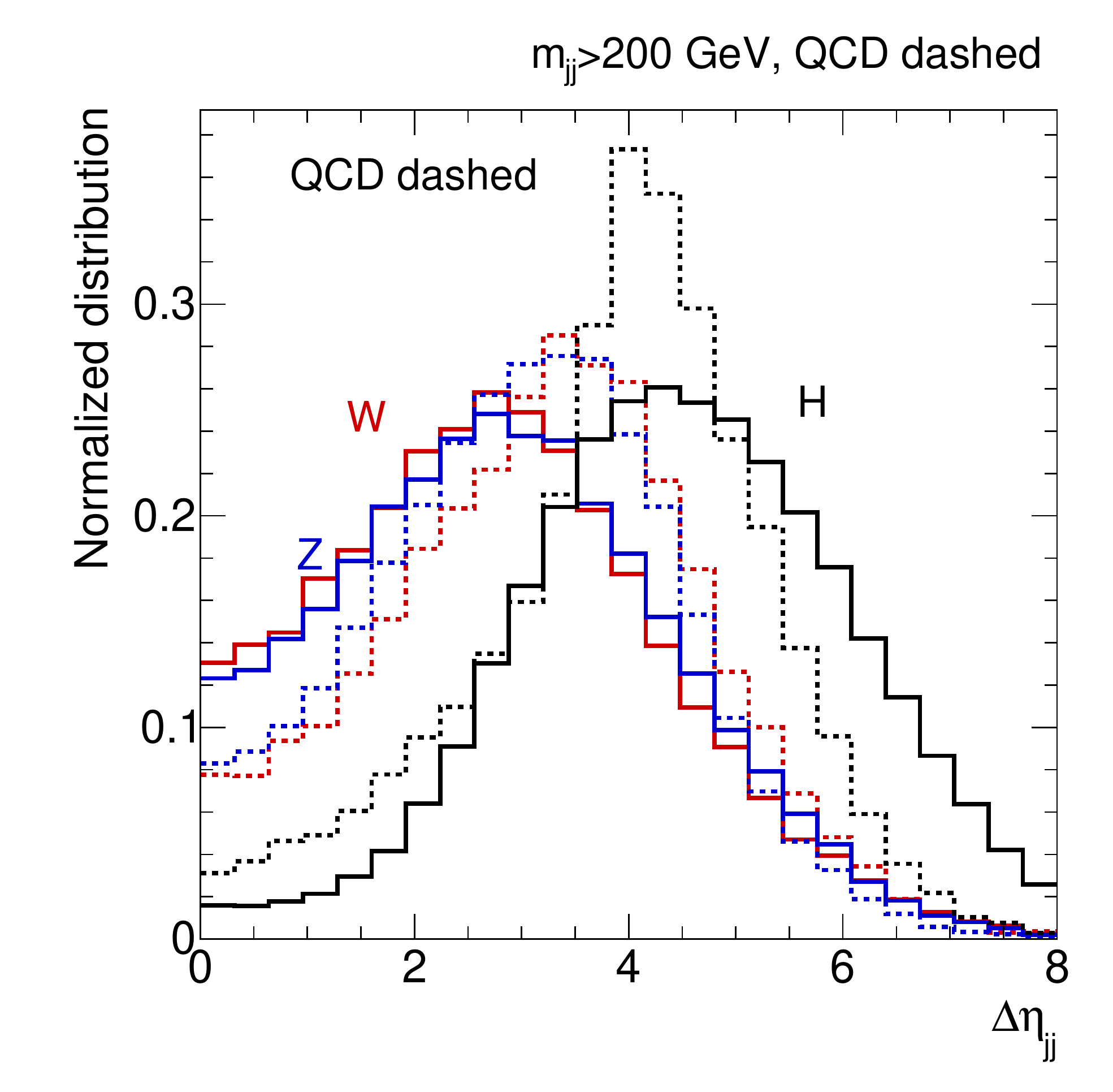}
\includegraphics[width=0.32\textwidth]{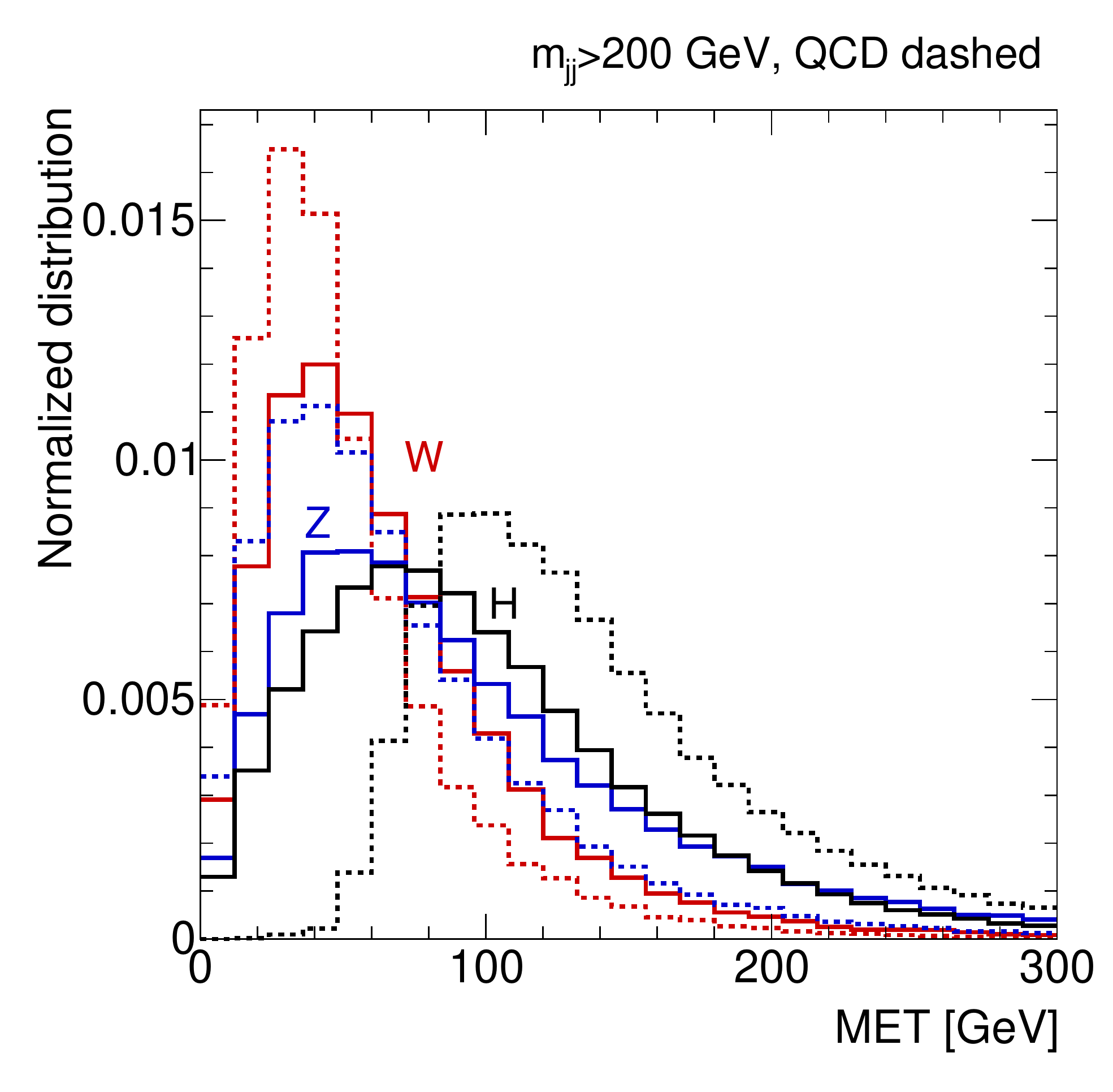}
\includegraphics[width=0.32\textwidth]{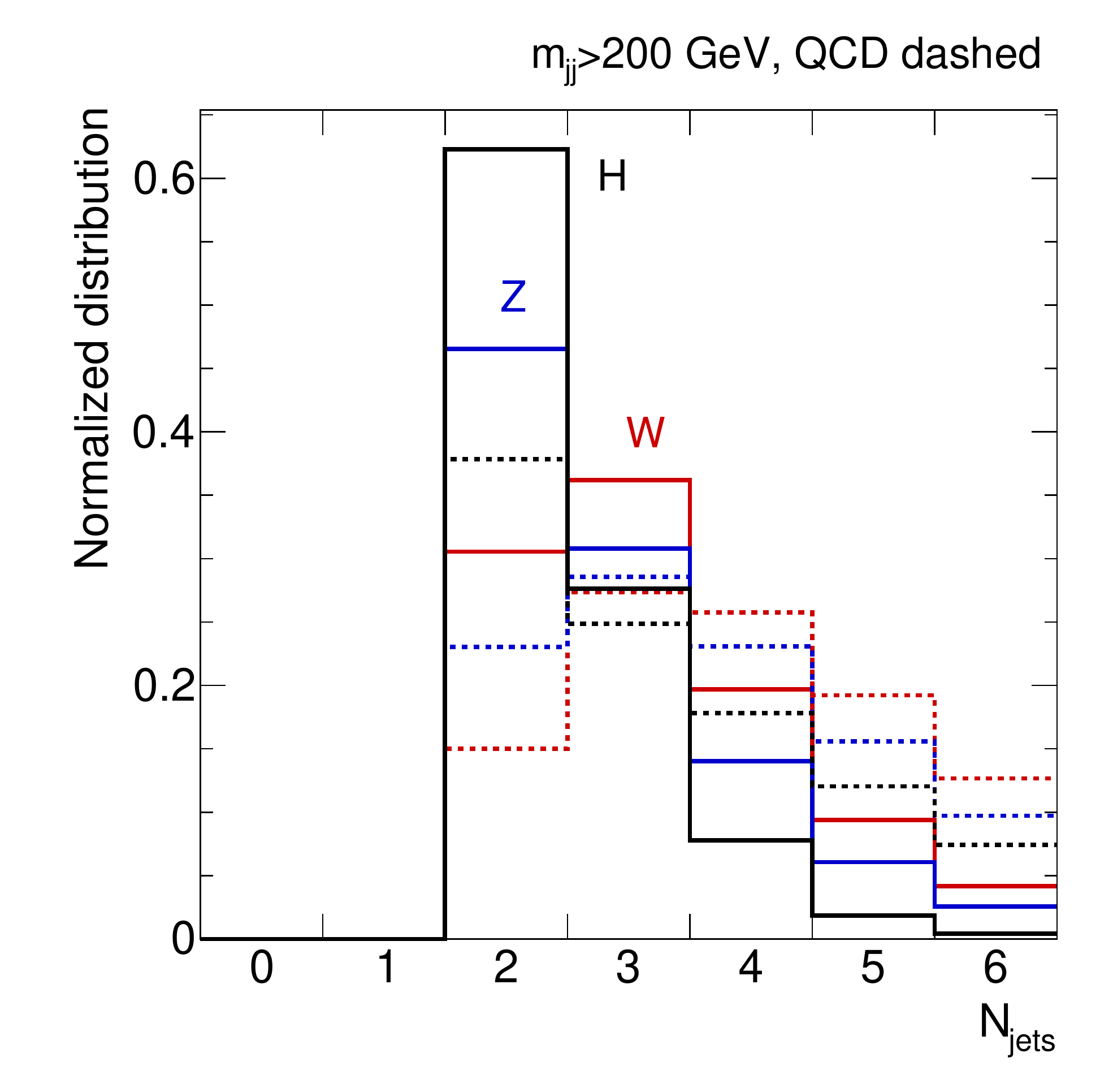} \\
\includegraphics[width=0.32\textwidth]{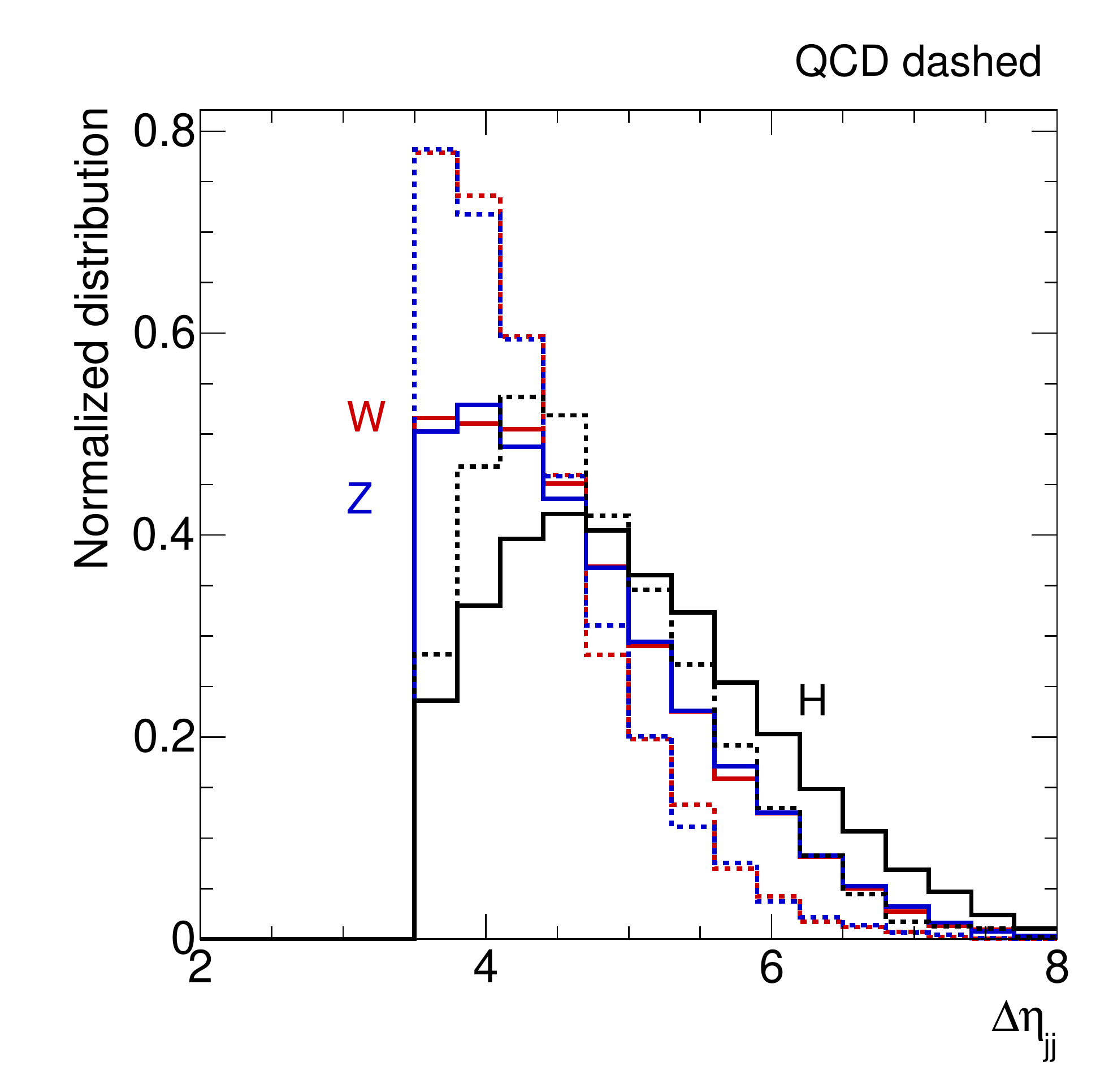}
\includegraphics[width=0.32\textwidth]{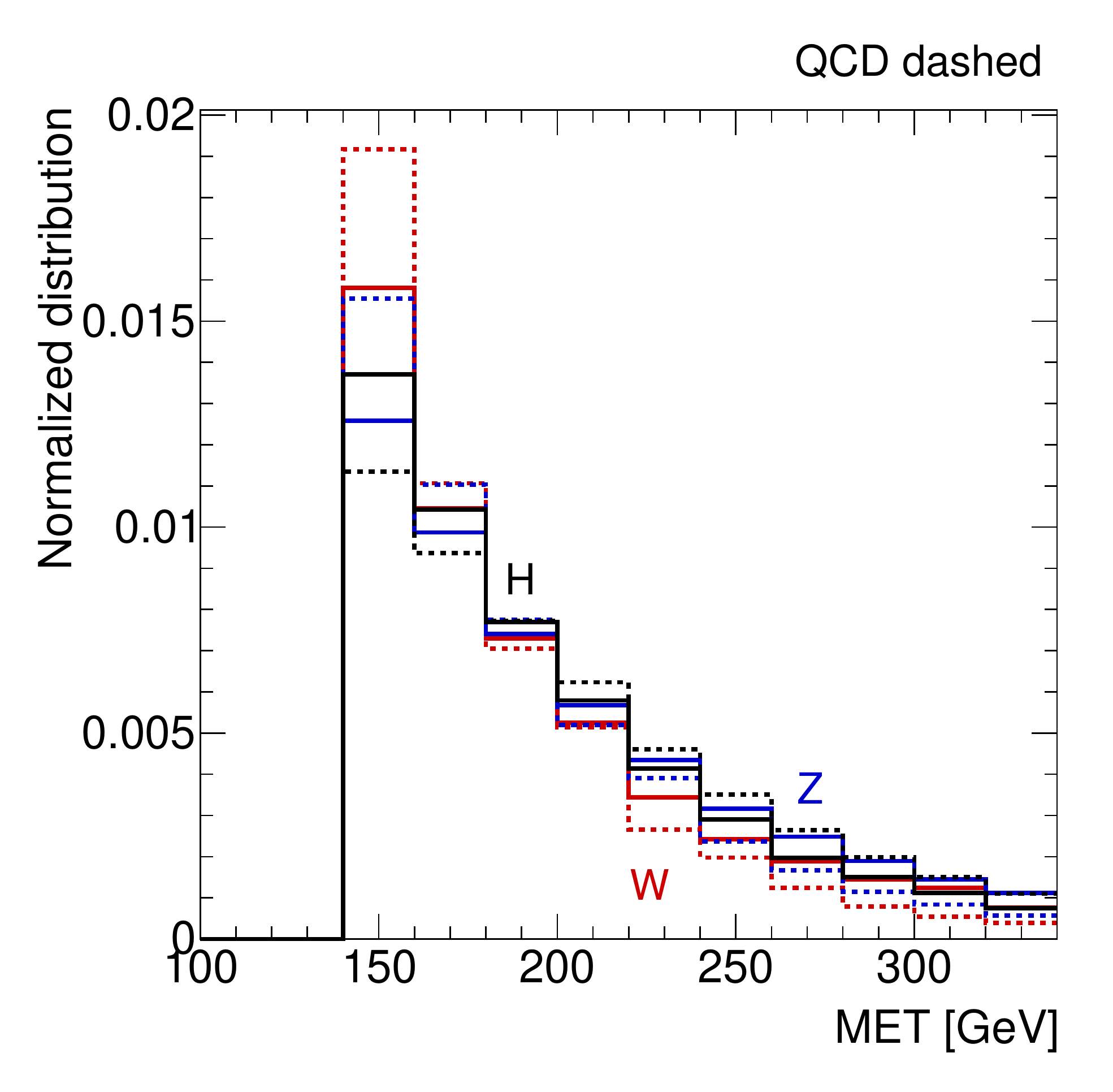}
\includegraphics[width=0.32\textwidth]{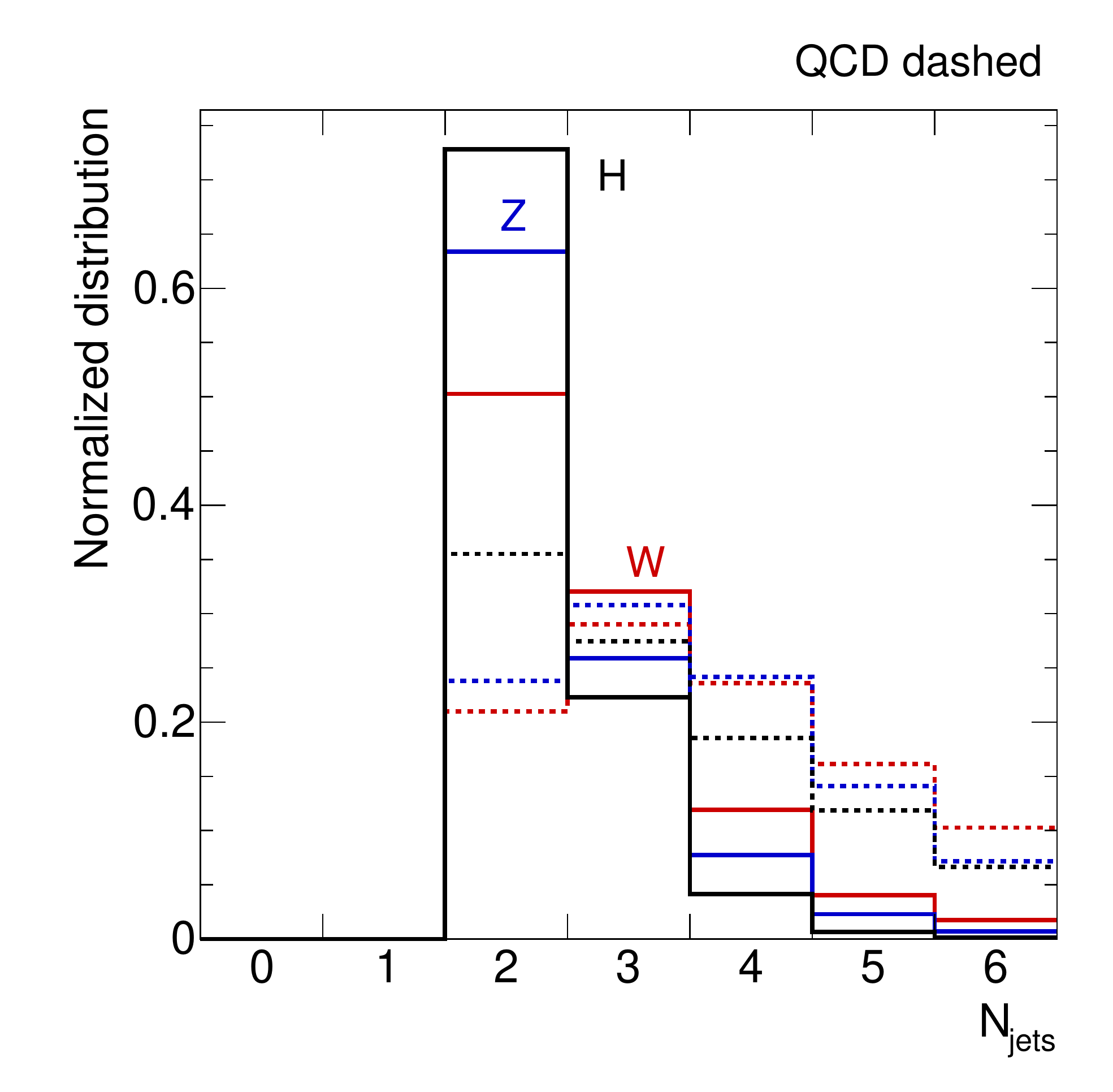}
\caption{WBF signal and background distributions after the minimal
  requirement $m_{jj}> 200$~GeV (upper panels) and after the
  pre-selection cuts of Eq.\eqref{eq:acc_cuts}. The $\Delta\eta_{jj}$
  (left) distribution for the $V$+jets $Z$ backgrounds look very
  similar. Deviations between $W$ and $Z$ backgrounds appear for the
  missing energy distribution (center) and $N_\text{jets}$ (right).}
\label{fig:wbf_missingLep}
\end{figure}

The leading channel to search for invisibly decaying Higgs bosons is
Higgs production in weak boson
fusion~\cite{eboli_zeppenfeld,spying}. The reason is that the recoil
against the tagging jets automatically leads to a transverse momentum
$p_{T,H} \gtrsim m_W/2$ and that the tagging jet structure with a
reduced central jet activity allows us to control QCD
backgrounds~\cite{taggingjets}. In principle, this channel could be
sensitive to invisible branching ratios down to $\br_\text{inv}
\approx 2\%~...~3\%$ for integrated luminosities of $3~\iab$. To test
branching ratios below the percent range we will have to go to a
100~TeV hadron collider~\cite{nimatron,spectroscopy}.

While the structure of the hard weak boson fusion process is
relatively simple, the LHC reach in this channel is largely determined
by our understanding of the backgrounds, including central-detector
QCD features in the signal and background events.  The main
irreducible background is $Z$+jets production with a decay $Z \to \nu
\bar{\nu}$. The nominally reducible $W$+jets production with an
unobserved decay lepton can contribute at a similar level.  The two
hard jets of the signal can be produced either through a hard QCD
process, $\sigma(Vjj) \propto \alpha_s^2 \alpha$, or through a hard
electroweak process $\sigma(Vjj) \propto \alpha^3$, with $V = Z,W$. It
is crucial to separate these two production mechanisms if we later use
the hadronic activity to separate the Higgs signal from those
backgrounds. 

The $Z$-background can be measured using the visible decays $Z \to
\ell \ell$, but in a control sample which due to the $Z$ branching
ratios is smaller than the actual background sample.  The $W$+jets
background with a leptonic decay $W \to \ell \bar{\nu}$ occurs when we
lose the lepton in the detector. Aside from the corresponding phase
space effects, preferring a forward or soft lepton, it should look
very similar to the $Z$+jets background.\bigskip

We simulate the WBF signal and its background processes for a 14~TeV
high-luminosity LHC at LO using \textsc{Sherpa2.2.1}~\cite{sherpa} with up to
three or four hard jets combined in the \textsc{Ckkw}
scheme~\cite{ckkw}. For the matrix element, we employ 
\textsc{Comix}~\cite{comix}. 
For both backgrounds we separate the QCD and weak
sub-processes.  For the signal we also take into account the
contribution from gluon fusion, denoted as QCD signal
contribution~\cite{ggFvsWBF}.  The corresponding event sample is
generated at LO with two hard jets using \textsc{Sherpa}, employing
\textsc{OpenLoops}~\cite{openloops, cuttools,oneloop,collier} to
include the finite top mass effects in the loop.  The
two tagging jet candidates are defined as the two hardest anti-$k_T$
jets in the event using \textsc{FastJet}~\cite{fastjet} with a jet
size $R = 0.4$. Detector effects are taken into account using
\textsc{Delphes3.3}~\cite{delphes} and the ATLAS card with an updated
lepton efficiency~\cite{atlas_elec_eff,atlas_muon_eff}.

In the upper panels of Fig.~\ref{fig:wbf_missingLep} we show the two
$W$ and $Z$ backgrounds in comparison to the Higgs signal after the
minimal requirement 
\begin{align}
m_{jj} > 200~\gev \; .
\label{eq:mjj200}
\end{align}
Because the signal process includes the $ZH \to (jj) H$ topology, we
need to apply this simple kinematic cut to select the WBF
diagrams.  For both the QCD and the weak sub-processes we observe
that the $\Delta\eta_{jj}$ (and $m_{jj}$) background distributions
look very similar for the $Z$+jets and $W$+jets backgrounds. The missing energy distribution is
significantly softer for the $W$+jets background, both in the QCD
case and in the electroweak case. Especially for the QCD process we do
not expect any difference between the $Z$+jets and $W$+jets
distributions before detector effects.  The QCD $W$+jets process only
contributes as a background if the lepton leaves sufficient tracks in
the detector to be reconstructed.  However, even though a lepton is
not reconstructed as such it will still deposit energy in the
calorimeter, leading to a reduced missing energy recoiling against the
visible constituents in the detector.  We have checked that our
observed difference in the $\met$ distribution is due to this effect
and easily accounted for in experiment.\bigskip

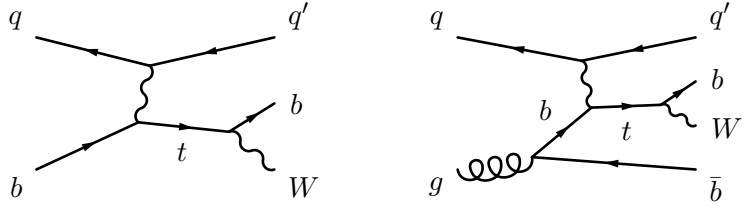
\begin{figure}[t]
\centering 
\begin{fmfgraph*}(90,50)
\fmfstraight
\fmfset{arrow_len}{2mm}
\fmfleft{i1,i2}  
\fmfright{o1,o2,o3}
\fmf{fermion}{o3,vu,i2}
\fmf{fermion}{i1,vd}
\fmf{fermion, lab=$t$}{vd,vW}
\fmf{fermion}{vW,o2}
\fmf{boson}{vu,vd}
\fmf{boson}{vW,o1}
\fmflabel{$b$}{i1}
\fmflabel{$q$}{i2}
\fmflabel{$W$}{o1}
\fmflabel{$b$}{o2}
\fmflabel{$q'$}{o3}
\end{fmfgraph*}
\qqqquad
\begin{fmfgraph*}(90,50)
\fmfstraight
\fmfset{arrow_len}{2mm}
\fmfleft{g1,i3}  
\fmfright{o1,o2,o3,o4} 
\fmf{fermion}{o4,vu,i3}
\fmf{fermion,lab.side=left}{o1,vd}
\fmf{fermion,lab=$b$,lab.side=left}{vd,vm}
\fmf{fermion,lab=$t$,lab.side=right}{vm,ve}
\fmf{fermion}{ve,o3}
\fmf{boson}{vm,vu}
\fmf{boson}{ve,o2}
\fmf{gluon,tension=3}{g1,vd} 
\fmflabel{$g$}{g1}
\fmflabel{$q$}{i3}
\fmflabel{$\bar{b}$}{o1}
\fmflabel{$W$}{o2}
\fmflabel{$b$}{o3}
\fmflabel{$q'$}{o4}
\end{fmfgraph*}
\caption{Example Feynman diagrams for the single top contribution to
  the $W$+jets background.}
\label{fig:feyn_njet_diff}
\end{figure}

The moment we apply something like a central jet
veto~\cite{eboli_zeppenfeld,spying}, the number of jets in the signal
and background events is a crucial observable.  We show the number of
jets with $p_{T,j} > 20$~GeV and $|\eta_j|< 4.5$ in the right panel of
Fig.\ref{fig:wbf_missingLep}. Instead of the expected similar behavior
for the $Z$+jets and $W$+jets backgrounds, we find significantly more
jets in electroweak $W$+jets events than in electroweak $Z$+jets
events.  Unlike the signal and all other backgrounds, the electroweak
$W$ background is more likely to include three or more jets in an
event.  The dominant effect that leads to this behavior is the
contribution of single top process to this background category,
\begin{align}
pp \to bW^+ \text{+ jets.}
\end{align}
We display the corresponding Feynman diagrams in
Fig.~\ref{fig:feyn_njet_diff}. For events with exactly two jets in the
final state, the contribution from those diagrams is $30\%$ for
$m_{jj} > 200$~GeV.  For three-jet events the single top process
can contribute up to $50\%$ of all events, eventually forcing us to
understand this difficult process with high precision during the
future LHC runs.\bigskip

Before we enter any dedicated signal vs background analysis we need to
apply a set of basic cuts, motivated by detector performance,
triggering, and rejecting generic backgrounds. Following the CMS
pre-selection~\cite{cms_comb_7813} we require two tagging jets combined
with sizeable missing transverse energy pre-selection
\begin{alignat}{7}
p_{T,j_{1,2}} &> 40 \, \gev  &\qqqquad
|\eta_{j_{1,2}}| &< 4.5 &\qqqquad 
\met &> 140 \, \gev \notag \\
\eta_{j_1} \; \eta_{j_2} &< 0 &
|\Delta \eta_{jj} | &> 3.5 &
m_{jj} &> 600 \, \gev \notag \\
p_{T,j_3} &> 20 \, \gev  &\qqqquad
|\eta_{j_3}| &< 4.5 &\qqqquad 
& \text{(if 3rd jet available)} \; .
\label{eq:acc_cuts}
\end{alignat}
and veto any additional observed lepton with a transverse momentum
larger than $p_{T,\ell}>7$~GeV based on our fast \textsc{Delphes3.3}
detector simulation.  In the lower panels of
Fig.~\ref{fig:wbf_missingLep} we show the corresponding distributions.
The good news is that once we apply these pre-selection cuts the
single top contamination in the $W$+jets sample drops to below $5 \%$
for two-jet events and $12\%$ for three-jet events. Therefore, the QCD
$V$+jets background resemble each other much more closely.

\section{Tagging jet size}
\label{sec:wbf_jetsize}

\begin{figure}[t]
\includegraphics[width=0.32\textwidth]{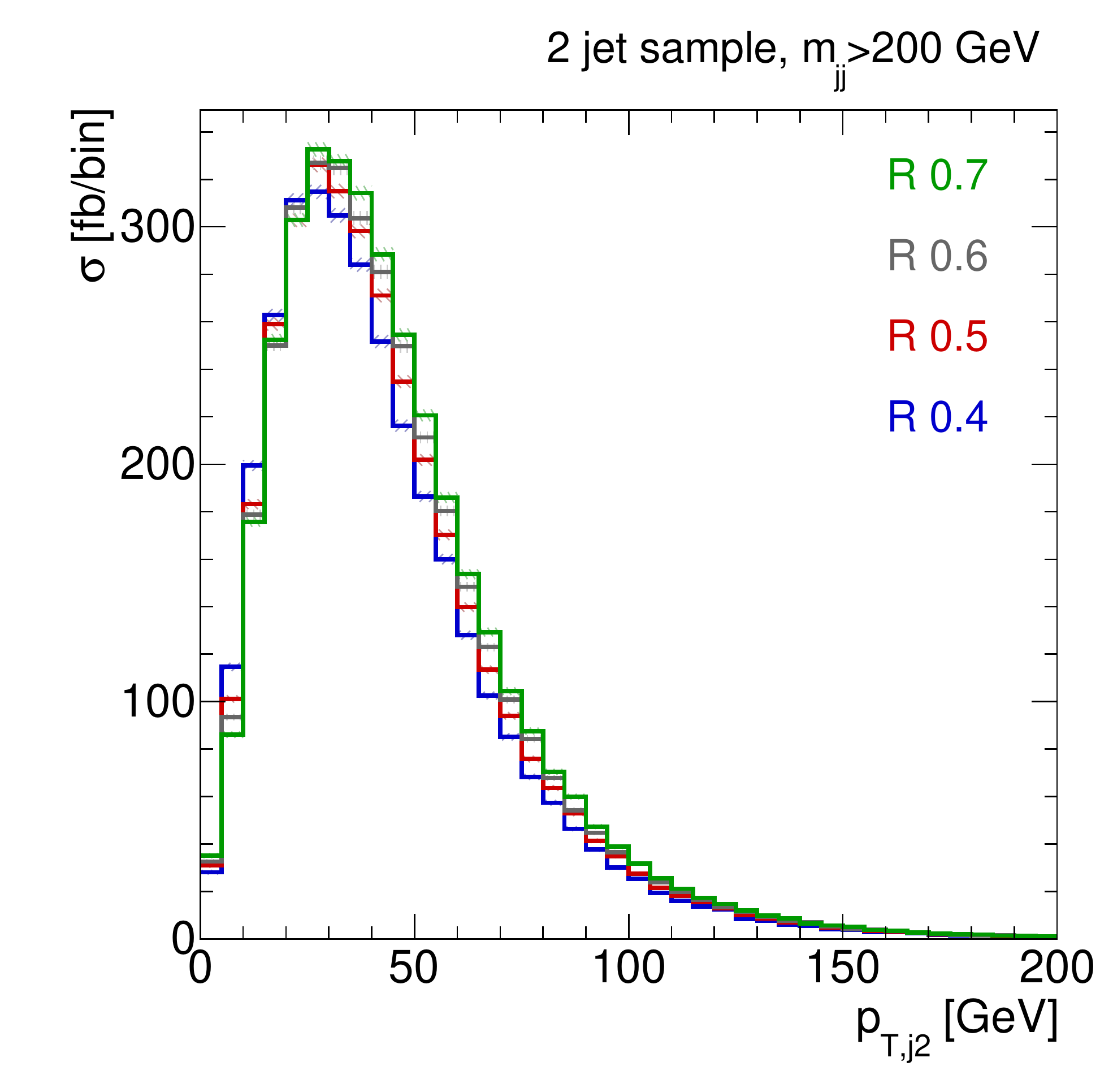}
\includegraphics[width=0.32\textwidth]{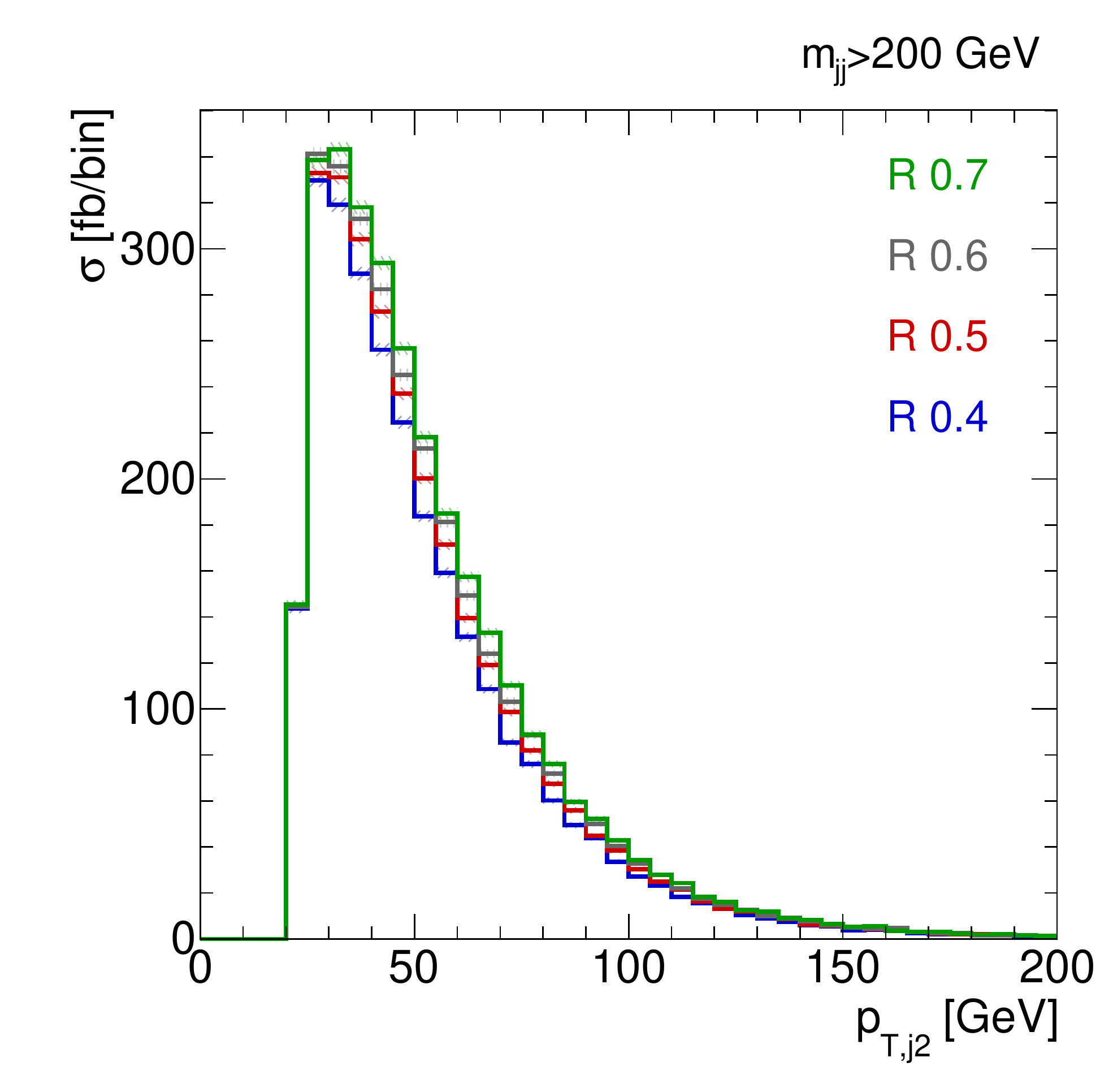}
\includegraphics[width=0.32\textwidth]{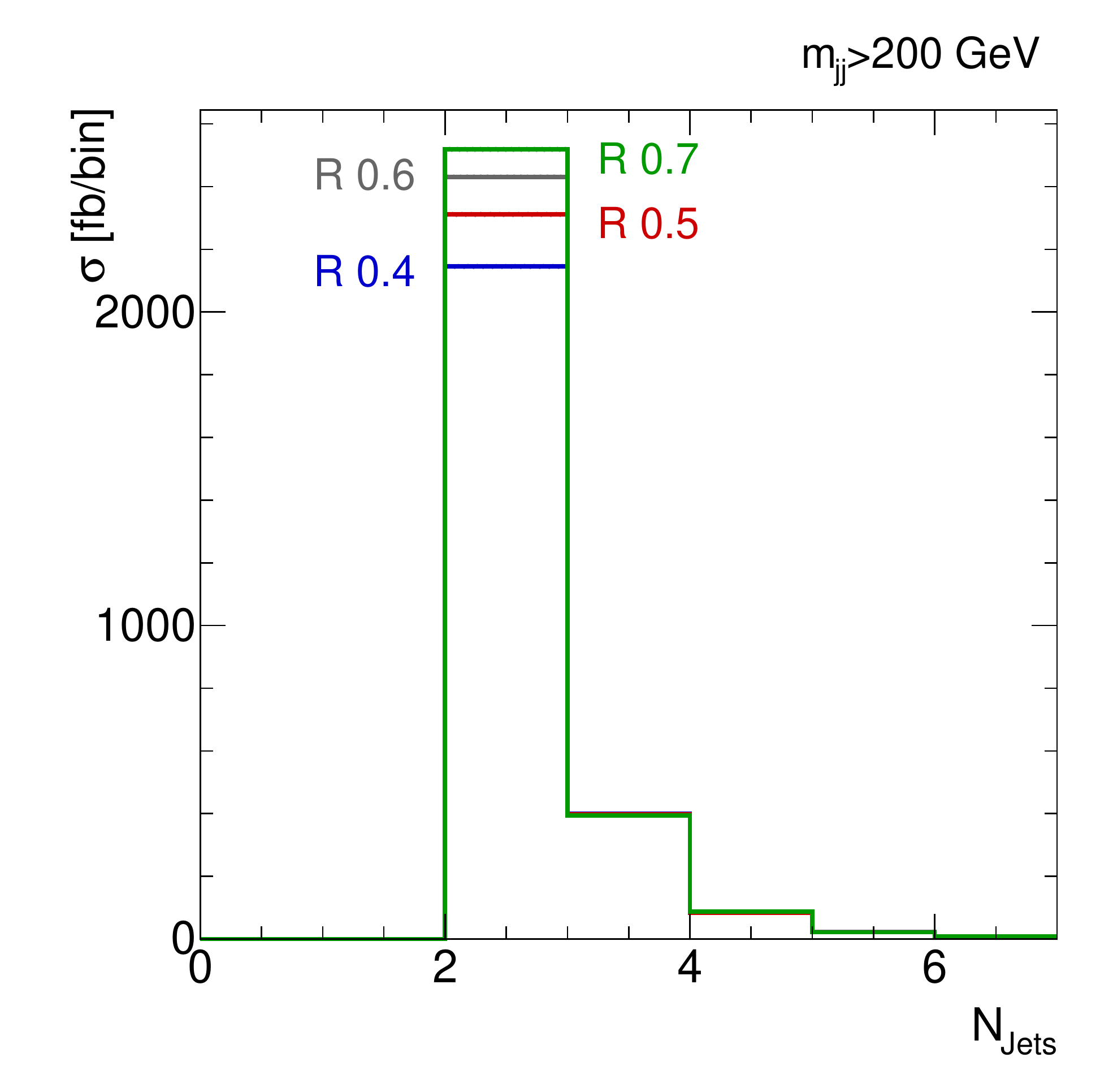}
\caption{WBF signal distributions for the exclusive two-jet sample
  (left) and the full sample (center and right) requiring only
  $m_{jj}> 200$~GeV.}
\label{fig:deltaR_dependence}
\end{figure}

Traditionally, the jet size in a given ATLAS or CMS analysis is chosen
following experimental considerations. However, as has been recently pointed
out, the geometric size of the tagging jets in WBF Higgs production
can have a sizeable impact on the effect of higher-order corrections
to the rate~\cite{mrauch}. Obviously, this is an effect of real parton
emission and how partons in addition to the two hard tagging jets are
combined into jets. For our simulations we use the same tool chain as
in the previous Sec.\ref{sec:wbf_irred}, combining up to three hard
jets and parton shower radiation.  We again define the tagging jets as
the hardest two jets with $p_{T,j} > 20$~GeV using the anti-$k_T$
algorithm in \textsc{FastJet}, but now with a variable jet size $R =
0.4~...~1.0$.\bigskip

We first show the $R$-dependence for the WBF signal 
\begin{align}
pp \to jj \; H_\text{inv}
\end{align}
in Fig.~\ref{fig:deltaR_dependence}. This signal
definition includes the $ZH$ topology, so we extract the WBF channel
in a two-jet plus parton shower setup with no $p_{T,j}$ cut, but
requiring $m_{jj}>200$~GeV as in Eq.\eqref{eq:mjj200}.  For the different
jet sizes we indeed find different $p_{T,j}$ spectra, for example for
the second tagging jet. The larger we choose the tagging jet, the
higher its transverse momentum becomes. The reason for this is simply
that the tagging jet is more likely to pick up additional jet
radiation, which by definition always increases its transverse
momentum. Between $R=0.4$ and $R=0.7$ the peak of the $p_{T,j}$
spectrum shifts by around 5~GeV. 
Including a third merged hard jet
reproduces this feature. Correspondingly, we also see that the number
of events with two jets increases with the size of the tagging
jets. Also the entire rate after the basic cut $m_{jj}>200$~GeV
increases, because the larger jet size helps the tagging jets to
collect hadronic activity and pass this minimal $m_{jj}$
requirement. This (accidentally) leaves the number of three-jet and
four-jet events constant.

\begin{figure}[b!]
\includegraphics[width=0.45\textwidth]{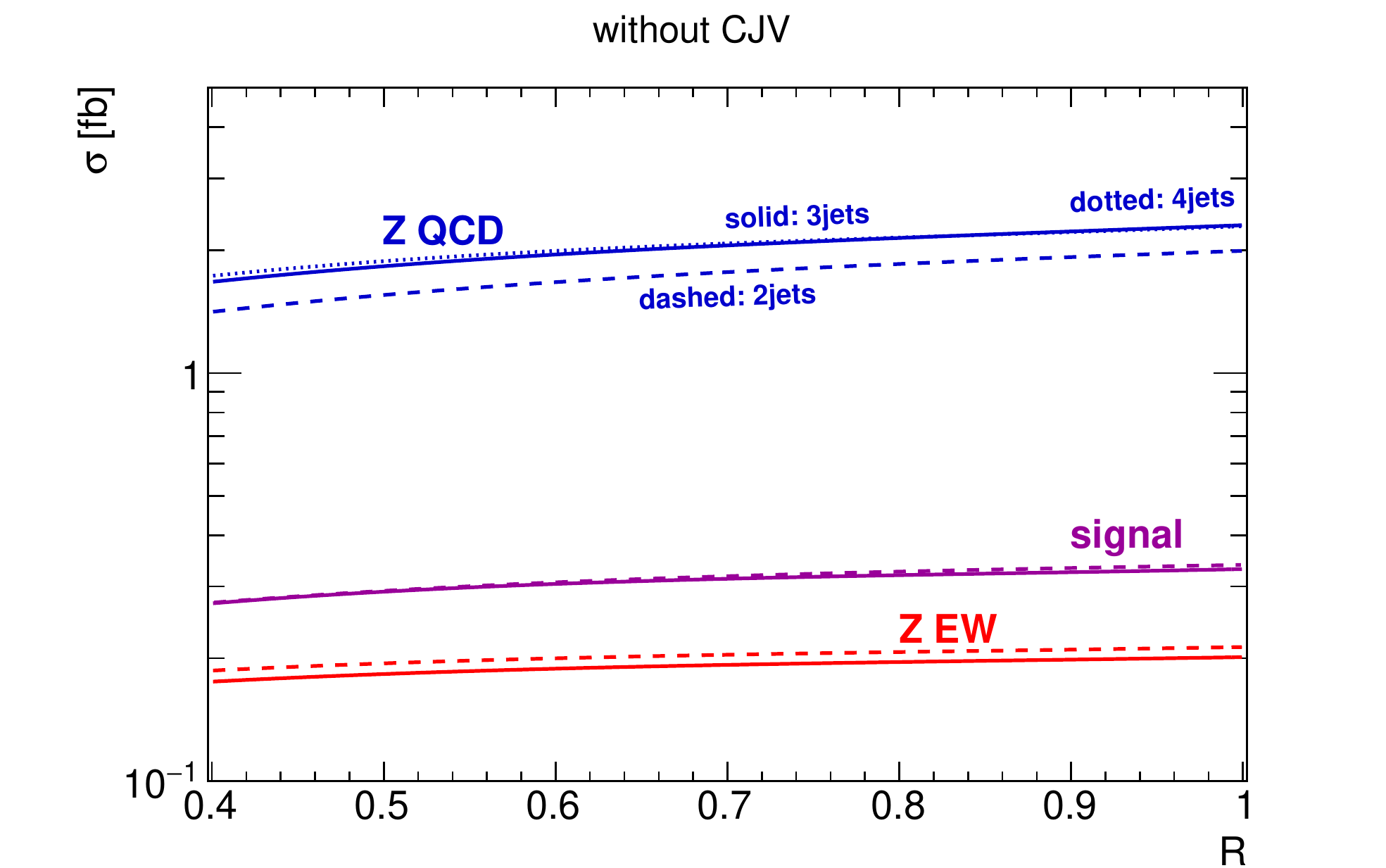} 
\hspace*{0.05\textwidth}
\includegraphics[width=0.45\textwidth]{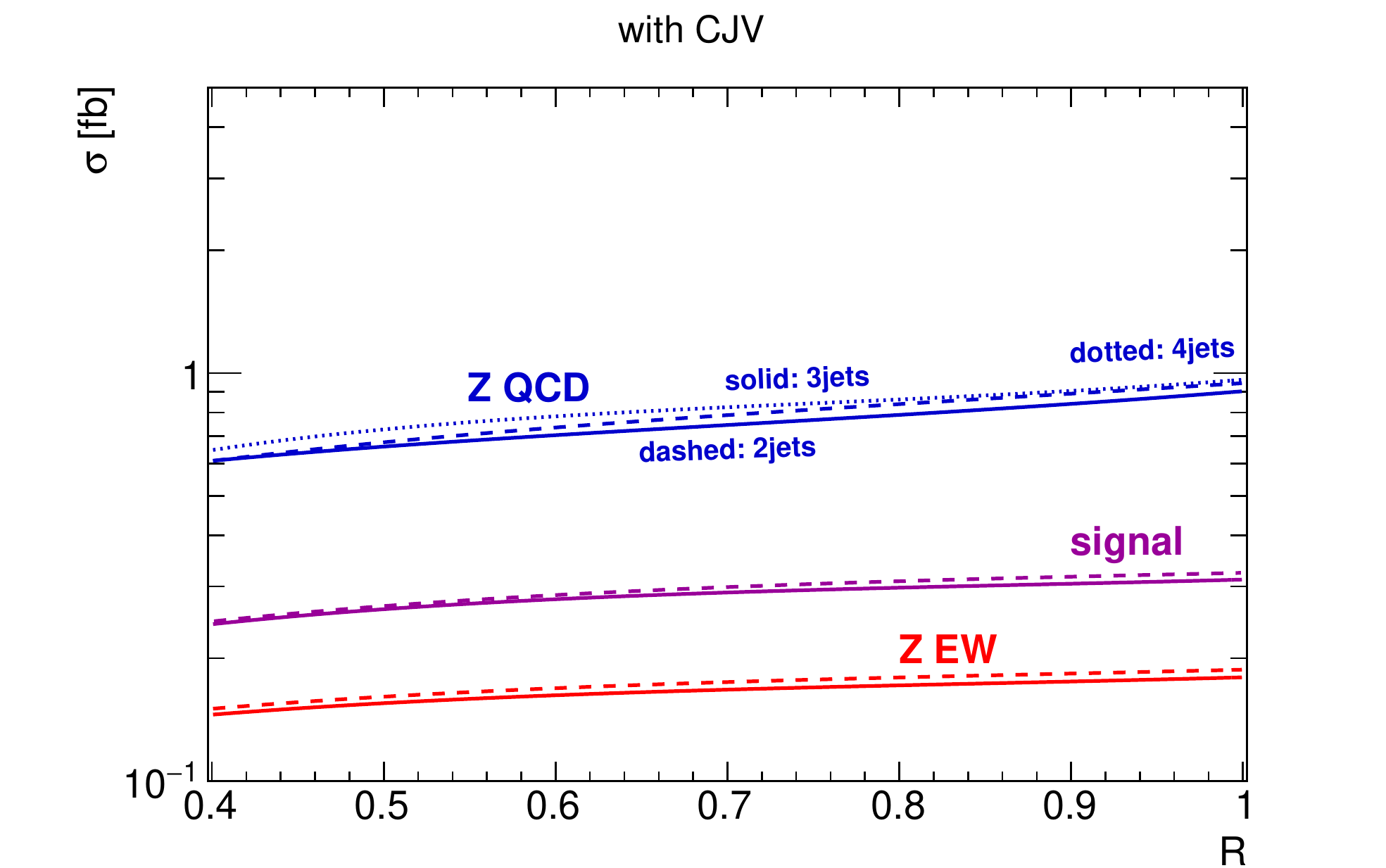} 
\caption{WBF signal and $Z$ background cross section dependence on $R$
  without (left) and with (right) central jet veto.  We
  always require the basic acceptance cuts of Eq.\eqref{eq:acc_cuts}.}
\label{fig:deltaR_1}
\end{figure}

Next, we show the actual $R$-dependence of the WBF signal and
$Z$-background rates in the left panel of Fig.~\ref{fig:deltaR_1}. To
obtain results actually relevant for the analysis we require the
pre-selection cuts of Eq.~\eqref{eq:acc_cuts}. For the QCD $Z$+jets
backgrounds we also show a curve with up to four merged hard jets. The
different lines correspond to two, three, and for the QCD background
four hard jets, merged and combined with the parton shower. As
expected, all rates increase with larger jet sizes. The question if we
simulate additional hard jets or generate these jets with the parton
shower plays essentially no role, independent of the nature of the
hard process.

In the left panel of Fig.~\ref{fig:deltaR_1} the QCD $Z$+jets process
shows a slightly larger slope than the electroweak signal and
background processes. This can be explained by the higher rate of jet
radiation off the QCD process and its external gluons.  We can force
the QCD and electroweak processes into more similar setups by
applying a central jet veto on jets with
\begin{align}
p_{T,j_3} > 20~\gev 
\qqqquad
\min \eta_{j_{1,2}}  < \eta_{j_3} < \max \eta_{j_{1,2}} \; .
\label{eq:cjv}
\end{align}
This way, the QCD $Z$+jets background will have a
reduced jet activity in the central detector. What is left is jet
radiation in the direction of the tagging jets.  In the right panel of
Fig.~\ref{fig:deltaR_1} we see the dramatic effect on the size of the
QCD background. Beyond this, the $R$-dependence of the different
signal and background processes is essentially identical.\bigskip

\begin{figure}[t]
\centering
\includegraphics[width=0.45\textwidth]{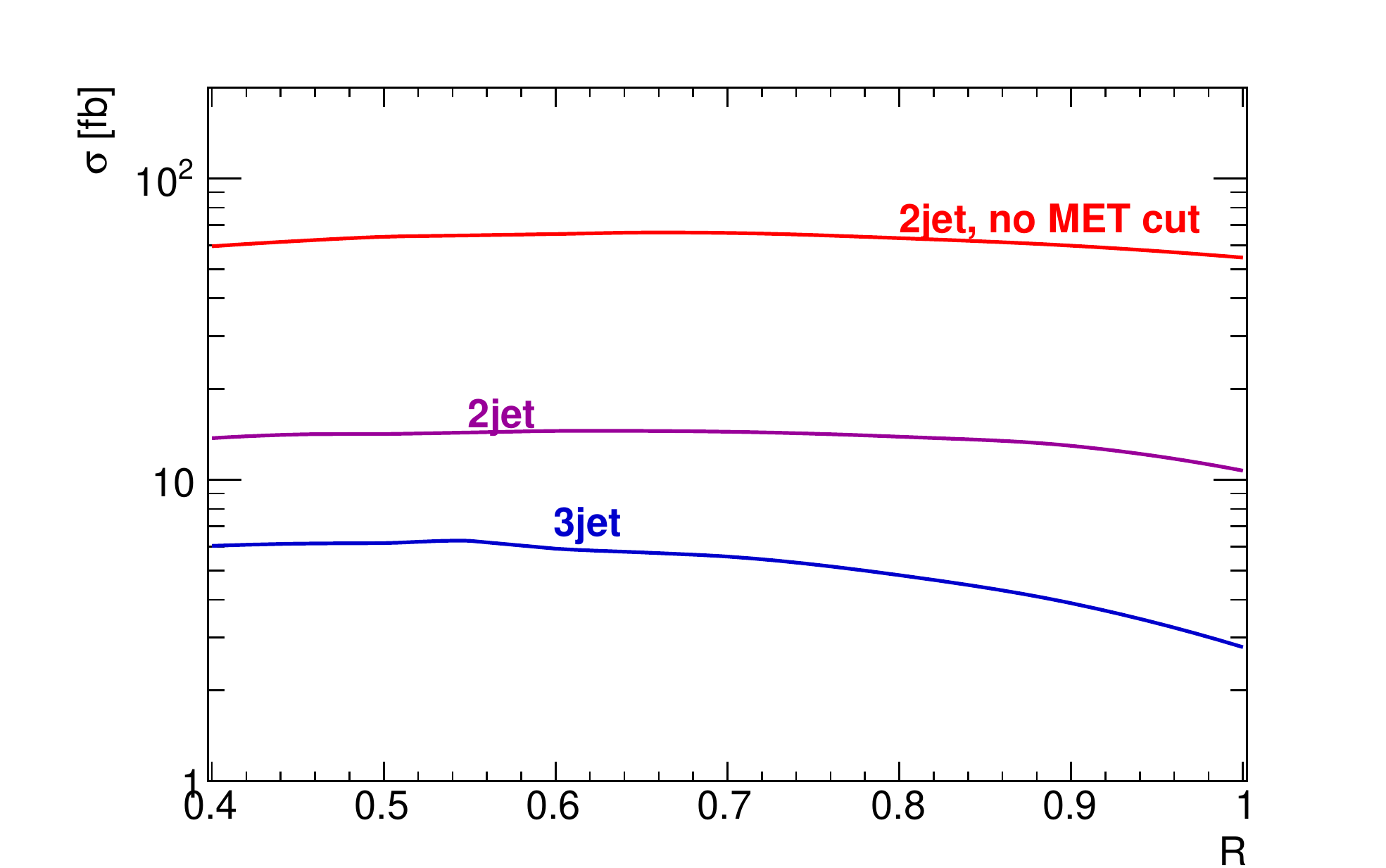}
\caption{Dependence of the $(Z \to jj)H$ rate on $R$ with the basic
  cuts of Eq.\eqref{eq:zh_zjj_cuts}.}
\label{fig:deltaR_2}
\end{figure}

Following this observation, the question arises whether the $R$-dependence
of the signal and background cross sections is a feature of the WBF
topology. As a test we analyze the process
\begin{align}
pp \to ZH_\text{inv} \to jj \; H_\text{inv} \, ,
\end{align}
which has the same final state as our signal, but a different topology. 
We again merge up to three jets and use \textsc{Delphes3.3} for detector simulation.
We apply the experimentally motivated cuts~\cite{atlas_zh}
\begin{alignat}{9}
N_\text{jets} &= 2,3 & \qqquad
p_{T,j_1} &> 45 \, \gev & \qqquad 
p_{T,j_{2,3}} &> 20 \, \gev \notag \\
\Delta R_{jj} &= 0.7~...~2.0 & \qqquad 
m_{jj}(2\,\text{jets}) &= 70~...~100 \, \gev & \qqquad 
m_{jj}(3\,\text{jets}) &= 50~...~100 \, \gev \notag \\
\met &= 120~...~160 \, \gev \; .
\label{eq:zh_zjj_cuts}
\end{alignat}
The difference between the $ZH$ topology and the WBF topology is that,
for the former, the two-jet system has a clear structure. First the
invariant mass of the two jets is fixed, and second the geometric
separation of the two jets can be related to the boost of the decaying
$Z$-boson.  In Fig.~\ref{fig:deltaR_2} we see that, for relatively
small jets, the rates for two-jet events and three-jet events are
stable. There are two features: first, the three-jet rate drops around
$R=0.55$. The corresponding topology is a boosted $Z$-boson recoiling
against the missing momentum and a third jet, where the two $Z$-decay
jets are boosted together into one observed jet. Second, the two-jet
sample shows a mild drop towards even larger jet sizes. In this case
the $Z$-boson recoils against the missing momentum alone and the point
at which the two decay jets are merged is moved to larger jet sizes,
$R$.  We confirm this pattern by looking at the two-jet sample without
the hard cut on $\met$, making it even less likely that the two
$Z$-decay jets are merged. Moreover, we observe that the relative
growth of the cross section for the two-jet sample without the hard
cut on $\met$ is comparable to the one for the WBF signal.

Altogether, our comparison shows that the observed $R$-dependence of
the WBF signal and background rates is not a specific feature of this
process, but instead an effect which appears more generally,
depending on the relevant phase-space regions. It is simply an effect
of extra jet radiation, and we observed a much more distinctive
$R$-dependence of the rate in the resonant $ZH$ topology.

\section{Tagging jet content}
\label{sec:wbf_jetcont}

After the pre-selection cuts of Eq.~\eqref{eq:acc_cuts} it turns out
that the largest backgrounds to invisible Higgs decays in WBF are clearly
QCD processes radiating a weak boson, $V = W,Z$.  Vetoing central
jets is an established, but tedious, way to reduce these QCD
backgrounds~\cite{eboli_zeppenfeld,spying}. However, a simple veto
is not necessarily the best use of the additional jet
information. This motivates a more comprehensive analysis of the jet
activity in the signal and the backgrounds~\cite{spying}.  In this
paper we go a step further and test how we can use observables linked
to quark vs gluon discrimination to control the QCD
backgrounds~\cite{quark_gluon,vikram}.\bigskip

\begin{figure}[t]
\includegraphics[width=0.32\textwidth]{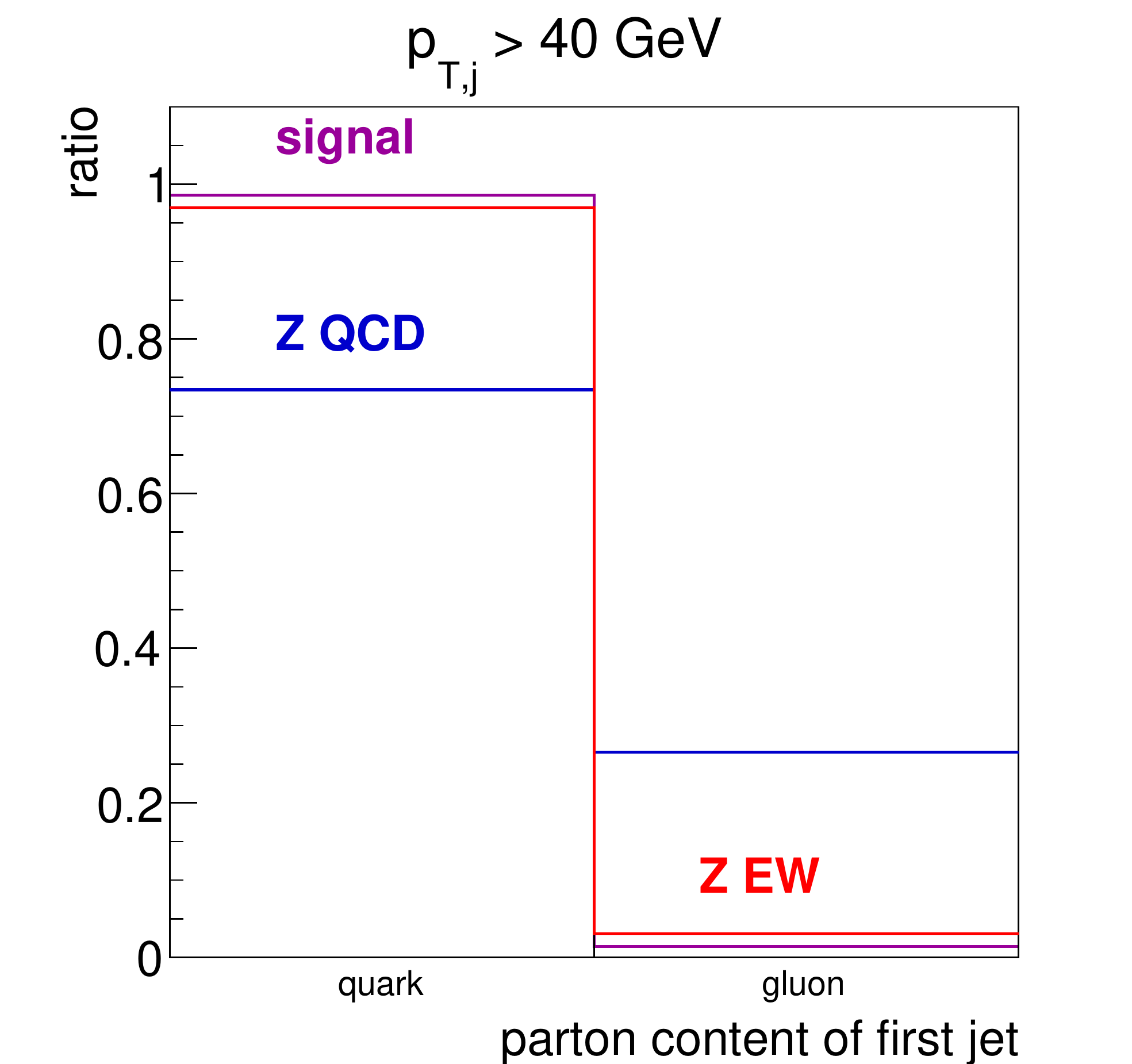}
\includegraphics[width=0.32\textwidth]{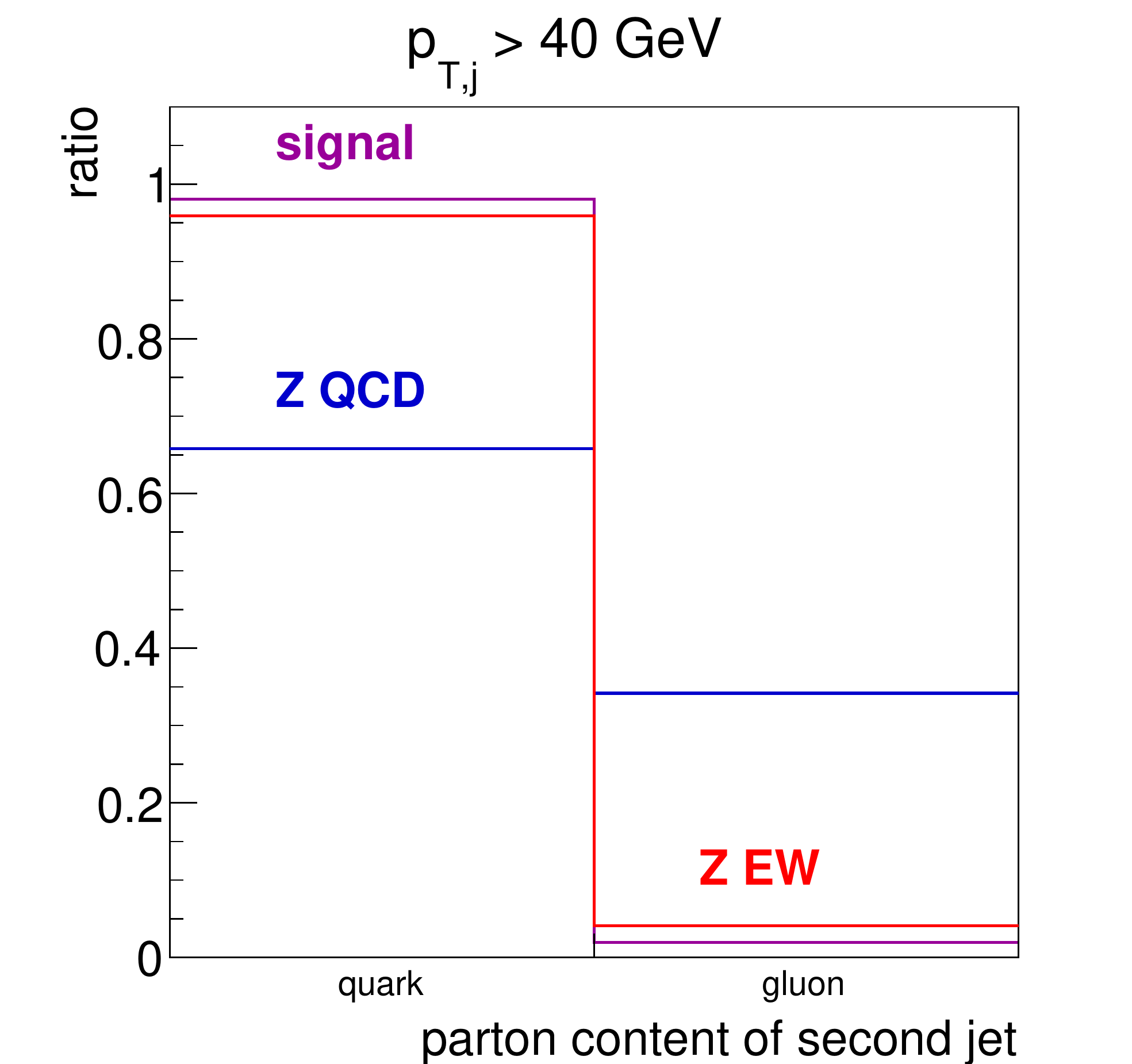}
\includegraphics[width=0.32\textwidth]{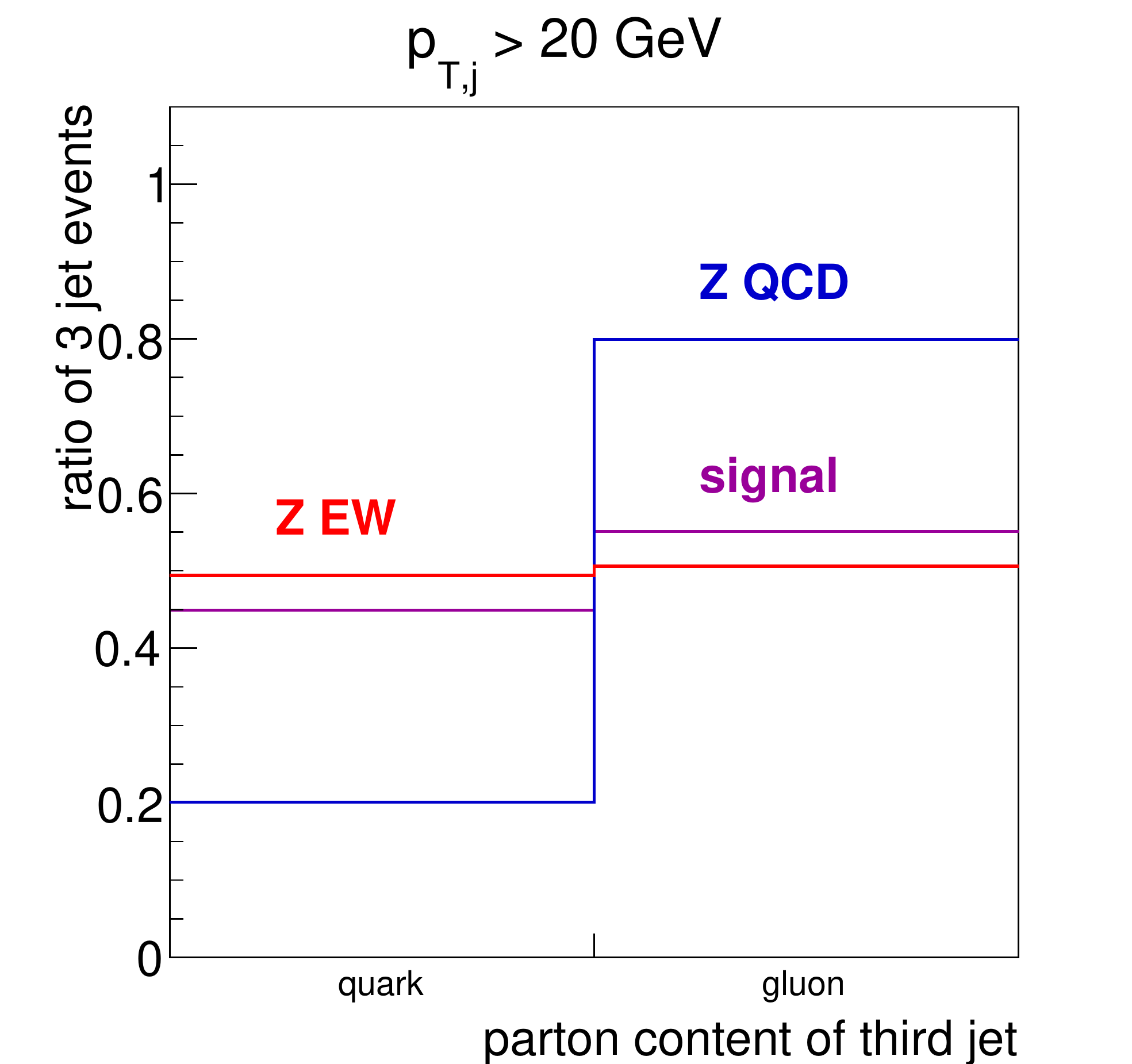}
\includegraphics[width=0.32\textwidth]{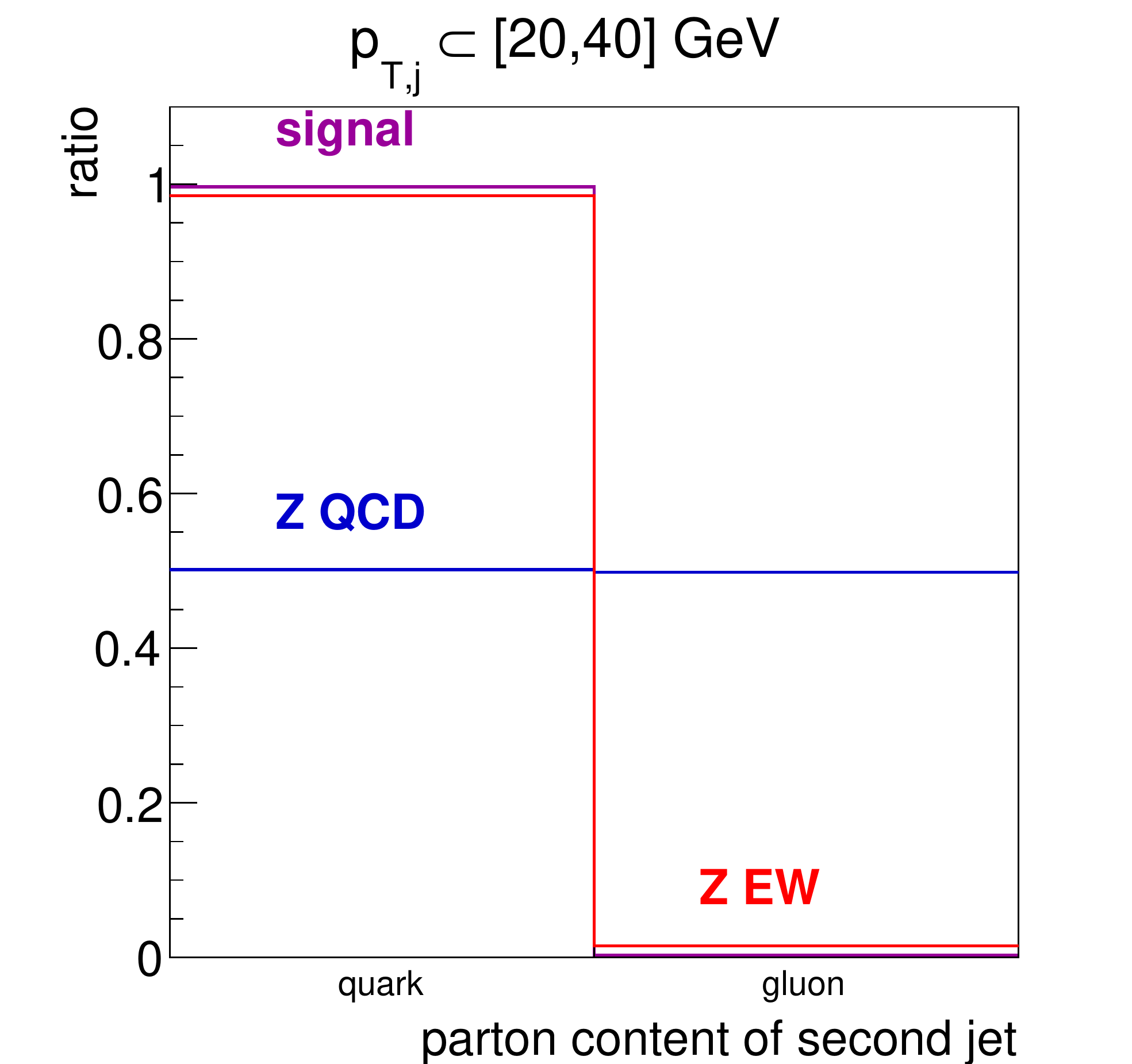}
\includegraphics[width=0.32\textwidth]{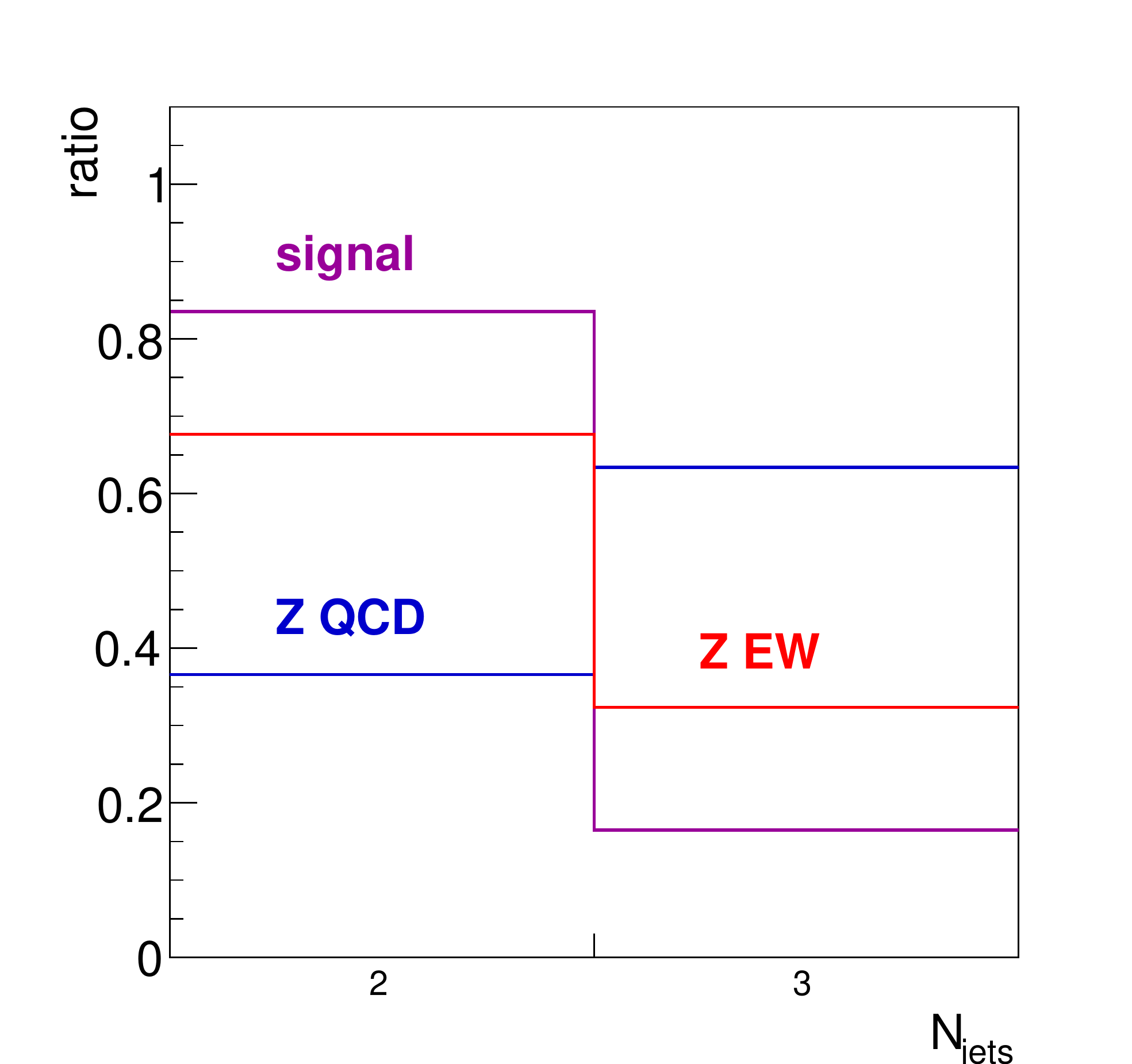}
\includegraphics[width=0.32\textwidth]{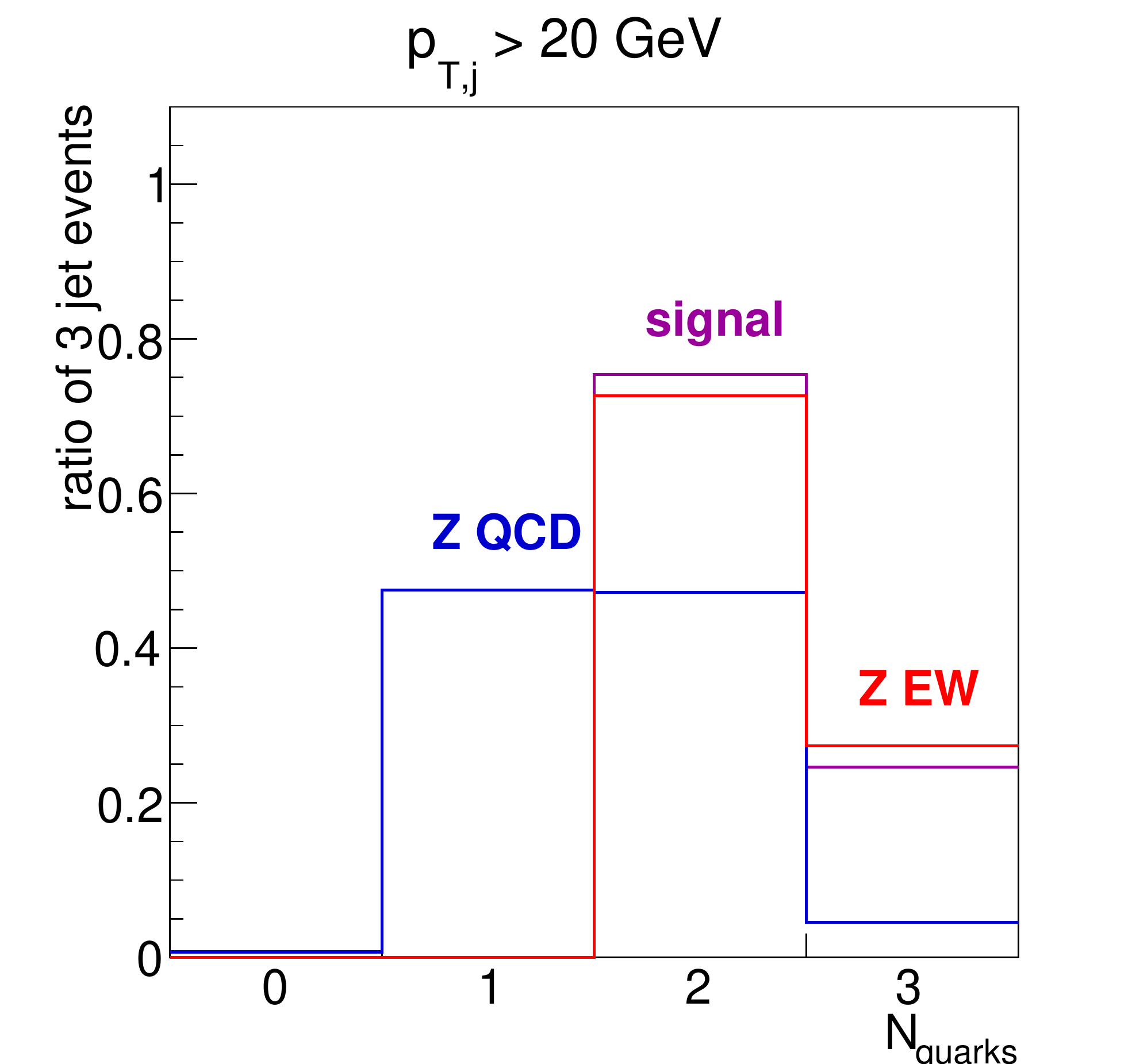}
\caption{Parton content of the first, second, and third jet for WBF
  signal and the backgrounds, after the pre-selection cuts of
  Eq.~\eqref{eq:acc_cuts}. In addition, we show the parton content of
  the second jet in the slice $p_{T,j_2} = 20~...~40$~GeV, the number
  of jets, and the number of quarks in three-jet events.}
\label{fig:qg_content_WBF}
\end{figure}

\begin{figure}[b!]
\centering 
\begin{fmfgraph*}(90,50)
\fmfstraight
\fmfset{arrow_len}{2mm}
	\fmfleft{in2,in1} 
	\fmfright{out2,outhiggs,out1} 
	\fmf{plain}{in1,vt}
	\fmf{plain}{vt,vup}
	\fmf{plain,tension=.5}{vup,out1}
	\fmf{boson}{vup,vinter}
	\fmf{boson}{vdown,vinter}
	\fmf{dashes}{vinter,outhiggs}
	\fmf{plain}{in2,vdown}
	\fmf{plain}{vdown,out2}
	\fmffreeze
	\fmftop{out3}
	\fmf{gluon}{out3,vt}
\end{fmfgraph*}
\qqqquad
\begin{fmfgraph*}(90,50)
\fmfstraight
\fmfset{arrow_len}{2mm}
	\fmfleft{in2,in1} 
	\fmfright{out2,outhiggs,out1} 
	\fmf{gluon}{in1,vt}
	\fmf{plain}{vt,vup}
	\fmf{plain,tension=.5}{vup,out1}
	\fmf{boson}{vup,vinter}
	\fmf{boson}{vdown,vinter}
	\fmf{dashes}{vinter,outhiggs}
	\fmf{plain}{in2,vdown}
	\fmf{plain}{vdown,out2}
	\fmffreeze
	\fmftop{out3}
	\fmf{plain}{vt,out3}
\end{fmfgraph*}
\qqqquad
\caption{Three-jet contribution to the WBF signal with two or three
  quarks in the final state.}
\label{fig:feyn_three_jet_signal}
\end{figure}
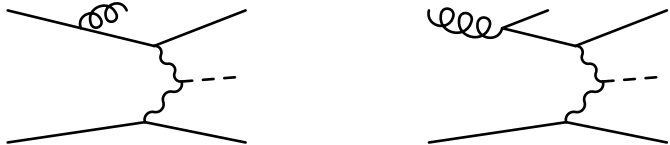

Before we apply quark-gluon discrimination to the WBF signal and the
QCD backgrounds we need to show that their partonic nature is indeed
significantly different to motivate an application of quark-gluon
discrimination. In general, we expect the tagging jets in the WBF
signal to almost exclusively be quark-initiated, as confirmed in
Fig.~\ref{fig:qg_content_WBF}.  For our parton-level illustration we
do not apply a detector simulation, but we apply the pre-selection
cuts of Eq.\eqref{eq:acc_cuts} with the exception of the $\met$
cut. This cut is strongly affected by the detector simulation, and we
replace it with a parton-level cut $p_T > 80$~GeV for the Higgs and
the $Z$-boson.  According to the pre-selection cuts, a third jet has
to be softer than the tagging jets. If the signal contains a third
jet, it is equally likely to be a quark and gluon. Around 50 $\%$ of
the events for which the third jet is a quark come from events with
two quarks and one gluon where the gluon is harder than at least one
of the quark jets, the other half is due to three quark events.  The
simple corresponding Feynman diagrams are given in
Fig.~\ref{fig:feyn_three_jet_signal}.

For the electroweak $Z$+jets background all panels in
Fig.~\ref{fig:qg_content_WBF} show essentially the same behavior as
for the signal, with a slightly larger probability to see a third jet,
reflecting the large number of topologies contributing to this
background~\cite{taggingjets}.

For the QCD $Z$+jets background the situation is completely different.
Here, at tree level the harder tagging jet arise from a gluon in
approximately 30\% of the events.  This fraction grows to around 35\%
for the second tagging jet and 80\% for a possible third jet. From
this estimate we expect the quark-gluon discrimination of the second jet
to be most promising. However, the parton content also depends on the
transverse momentum cut on the jet, as displayed in
Fig.~\ref{fig:qg_content_WBF}. Therefore, the discrimination power for
the third jet should benefit from the lower $p_T$ threshold of this
additional, central jet.\bigskip

\begin{figure}[t]
\includegraphics[width=0.32\textwidth]{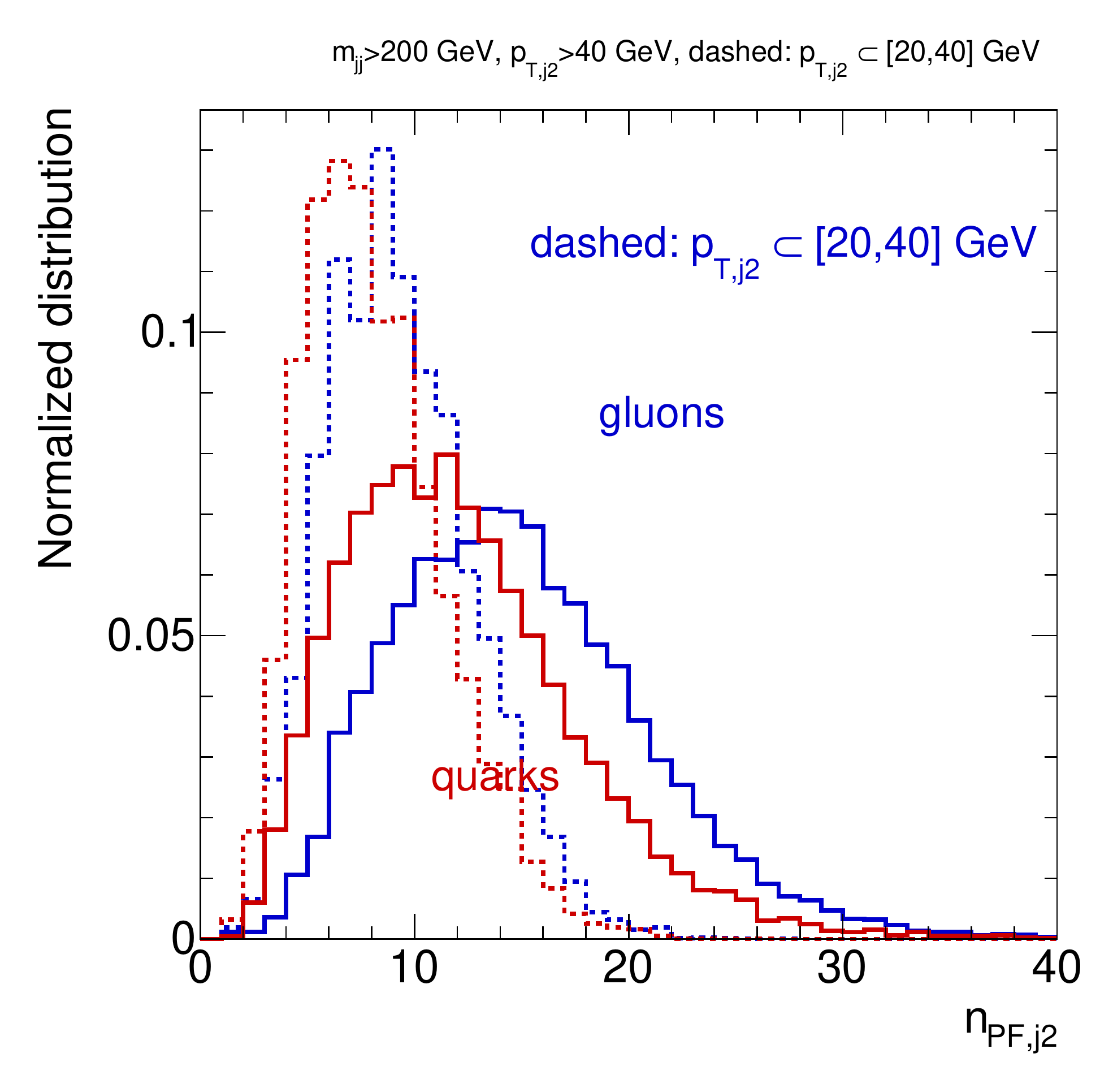}
\includegraphics[width=0.32\textwidth]{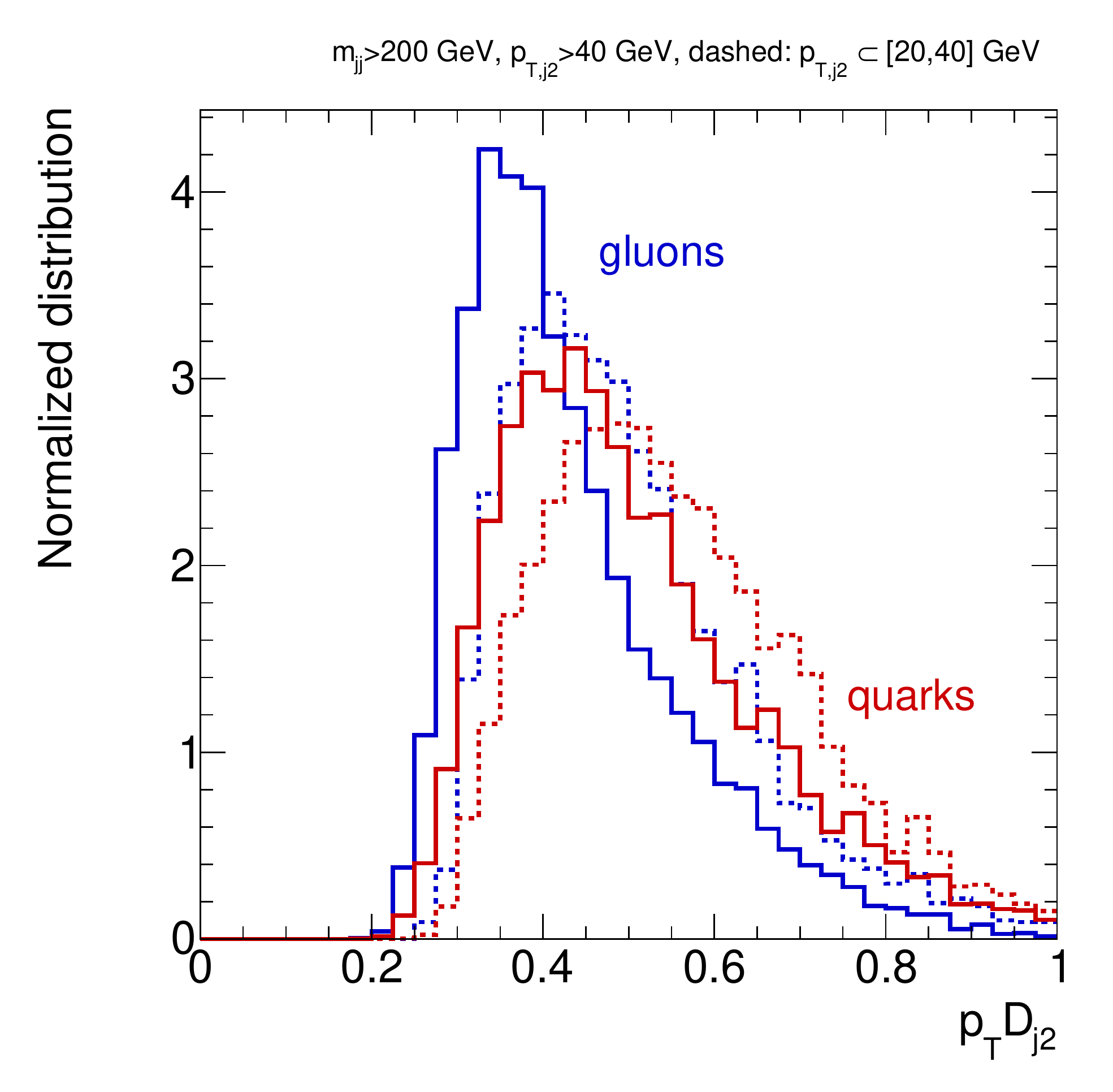}
\includegraphics[width=0.32\textwidth]{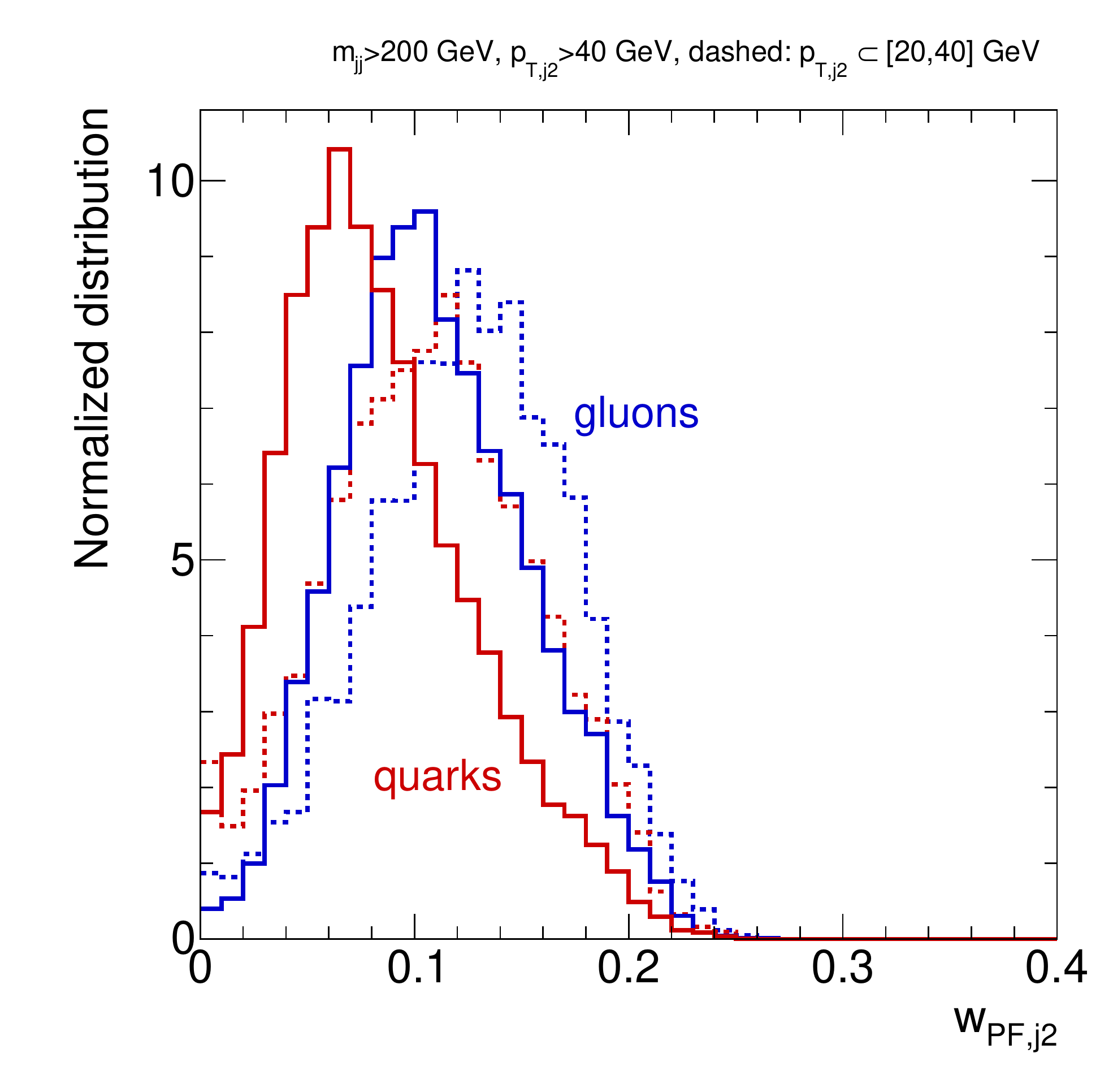} \\
\includegraphics[width=0.32\textwidth]{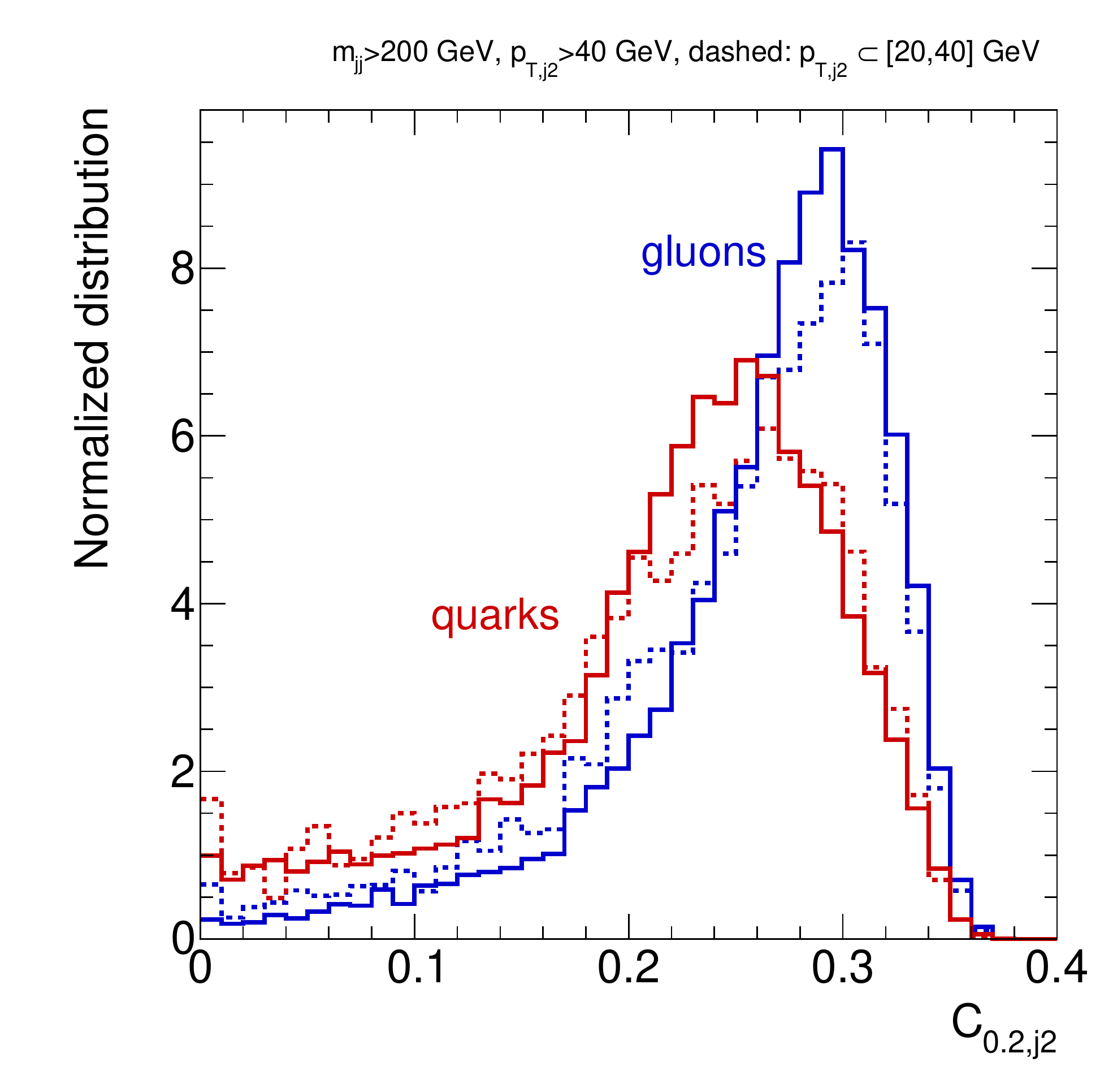}
\includegraphics[width=0.32\textwidth]{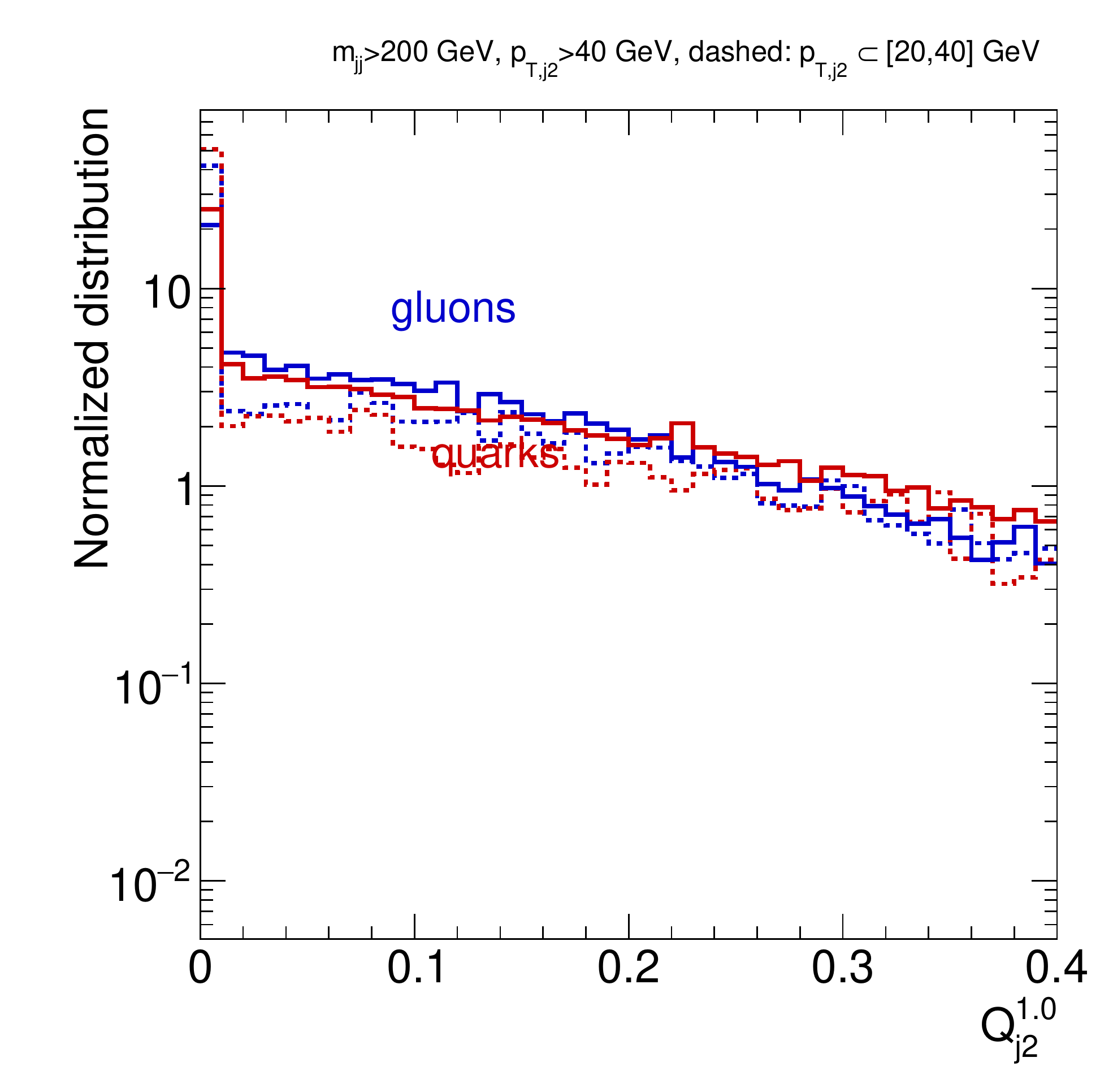}
\phantom{\includegraphics[width=0.32\textwidth]{figures/WBF14_Z_qcd_gq_j2qfrac}}
\caption{Distribution of the quark vs gluon discrimination variables
  listed in Eq.\eqref{eq:qg_obs} for the second jet in a pure QCD
  $Zjj$ sample. We show partonic two-quark and two-gluon final states
  requiring $m_{jj} > 200$~GeV combined with the two slices
  $p_{T,j_2} = 20~...~40$~GeV and $p_{T,j_2}>40$~GeV.}
\label{fig:wtrk_dist}
\end{figure}

After establishing that quark-gluon discrimination can be useful to
separate the WBF signal from QCD backgrounds we turn to appropriate
observables.  Some standard observables for quark vs gluon
discrimination can be easily expressed in terms of particle flow (PF)
objects or charged tracks as implemented in
\textsc{Delphes3.3}. They include~\cite{qg_dist:n90,qg_dist:wPF,qg_dist:C,qg_dist:Qk,qg_dist:pTD}
%
\begin{alignat}{9}
n_\text{PF} &= \sum_{i_\text{PF}} 1 &\qqqquad
C &= \frac{\sum_{i_\text{PF},j_\text{PF}} \; E_{T,i} E_{T,j} \; \left(\Delta R_{ij} \right)^{0.2}}
           {\left( \sum_{i_\text{PF}} E_{T,i} \right)^2} \notag \\
p_{T}D &= \frac{ \sqrt{ \sum_{i_\text{PF}} \; p_{T,i}^2 } }
          { \sum_{i_\text{PF}} p_{T,i} }  &\qqqquad 
Q^\kappa   &= \frac{\sum_{i_\text{trk}} \; q_i \; p_{T,i}^\kappa}
                  {\sum_{i_\text{trk}} \; p_{T,i}^\kappa} \notag \\
w_\text{PF} &= \frac{\sum_{i_\text{PF}} \; p_{T,i} \; \Delta R_{i, \text{jet}}}
                   {\sum_{i_\text{PF}} \; p_{T,i}} \; .
\label{eq:qg_obs}
\end{alignat}
We define all observables except for $Q^\kappa$ on particle flow
objects inside an anti-$k_T$ jet of size $R=0.4$.  In
Fig.~\ref{fig:wtrk_dist} we compare these for two idealized samples of
exclusive QCD $Zjj$ events, namely events with either two quarks or
two gluons in the final state. We apply a parton shower and the 
\textsc{Delphes3.3} detector simulation on the samples and 
only require the minimal selection
cut $m_{jj}>200$~GeV. We show the results for the second tagging jet in
two slices
\begin{align}
p_{T,j_2} = 20~...~40 \, \gev 
\qquad \text{and} \qquad 
p_{T,j_2} > 40 \, \gev \; .
\end{align}
We observe clear differences between the quark samples and gluon
samples in all distributions except for $Q^{1.0}$. Starting with
$n_{\text{PF},j_2}$, gluon jets clearly lead to more particle flow
objects because they are more likely to radiate or split and therefore
leave more tracks in the jet area.  In addition, harder jets generally
lead to more objects $n_{\text{PF},j_2}$.  The effect of the transverse
momentum on $n_{\text{PF},j_2}$ is significantly stronger than the
difference between quarks and gluons.  The variable $p_{T}D$ introduces
a transverse momentum weight to maintain infrared safety, but after
accounting for the inversion it follows a similar patterns as
$n_{\text{PF},j_2}$, with a reduced dependence on the transverse
momentum of the jet.

Adding the angular distance to the jet axis in $w_\text{PF}$ leaves
the quark vs gluon separation similar to $n_{\text{PF},j_2}$, but based
on pure kinematics the softer jets now systematically reside at lower
values of $w_\text{PF}$. Again expanding $w_\text{PF}$ to
object-object correlators in $C$ almost de-correlates the quark-gluon
discrimination from the transverse momentum of the jet. Finally,
$Q^{1.0}$ is clearly not expected to be very useful, because all
curves coincide for a large range of $Q^{1.0}$-values. Given the
similarity of the four leading distributions and the sizeable
correlation with the transverse momentum suggests a dedicated
multi-variate analysis to compare their performance.\bigskip

\begin{figure}[t]
\includegraphics[width=0.32\textwidth]{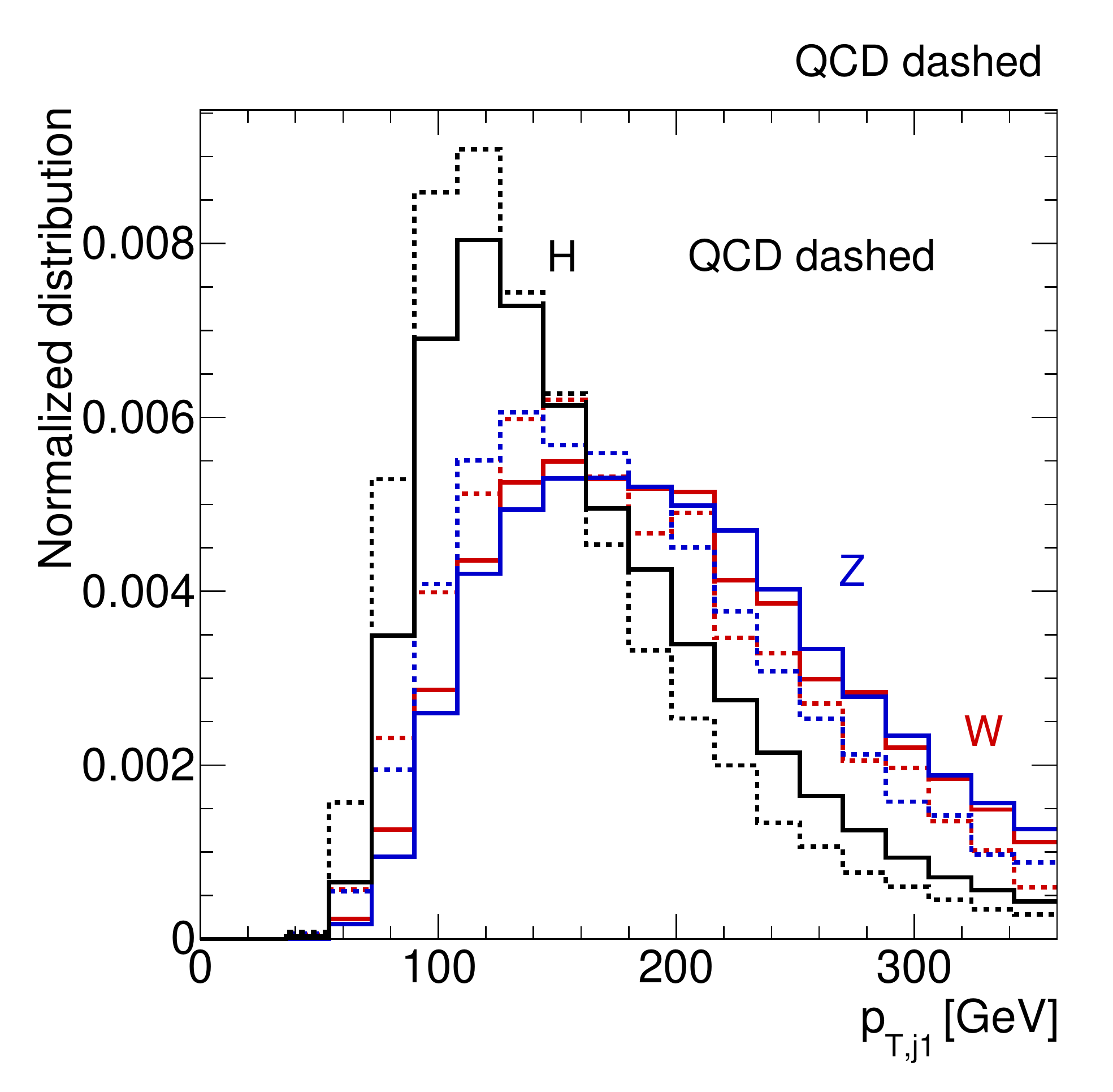}
\includegraphics[width=0.32\textwidth]{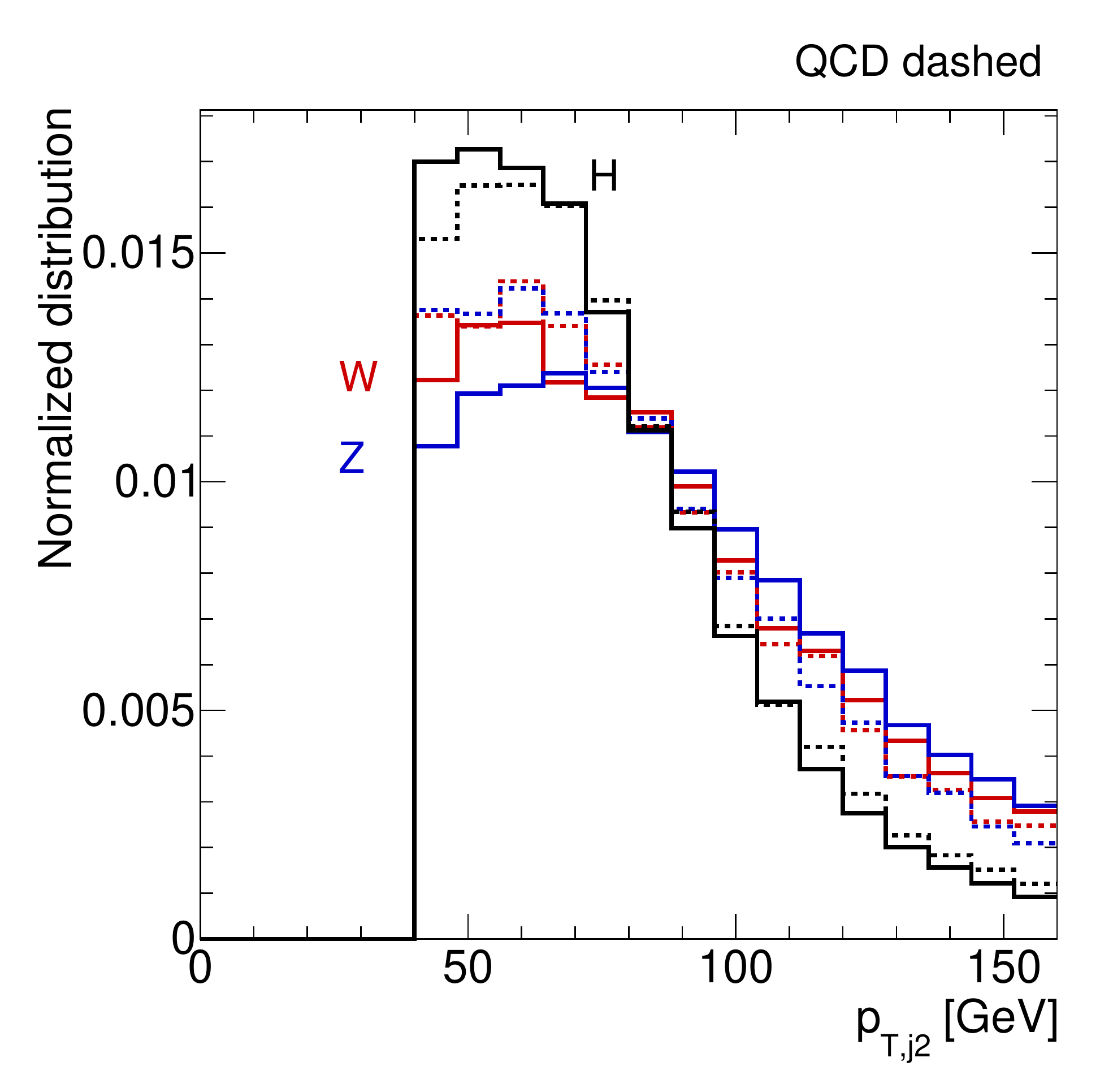}
\includegraphics[width=0.32\textwidth]{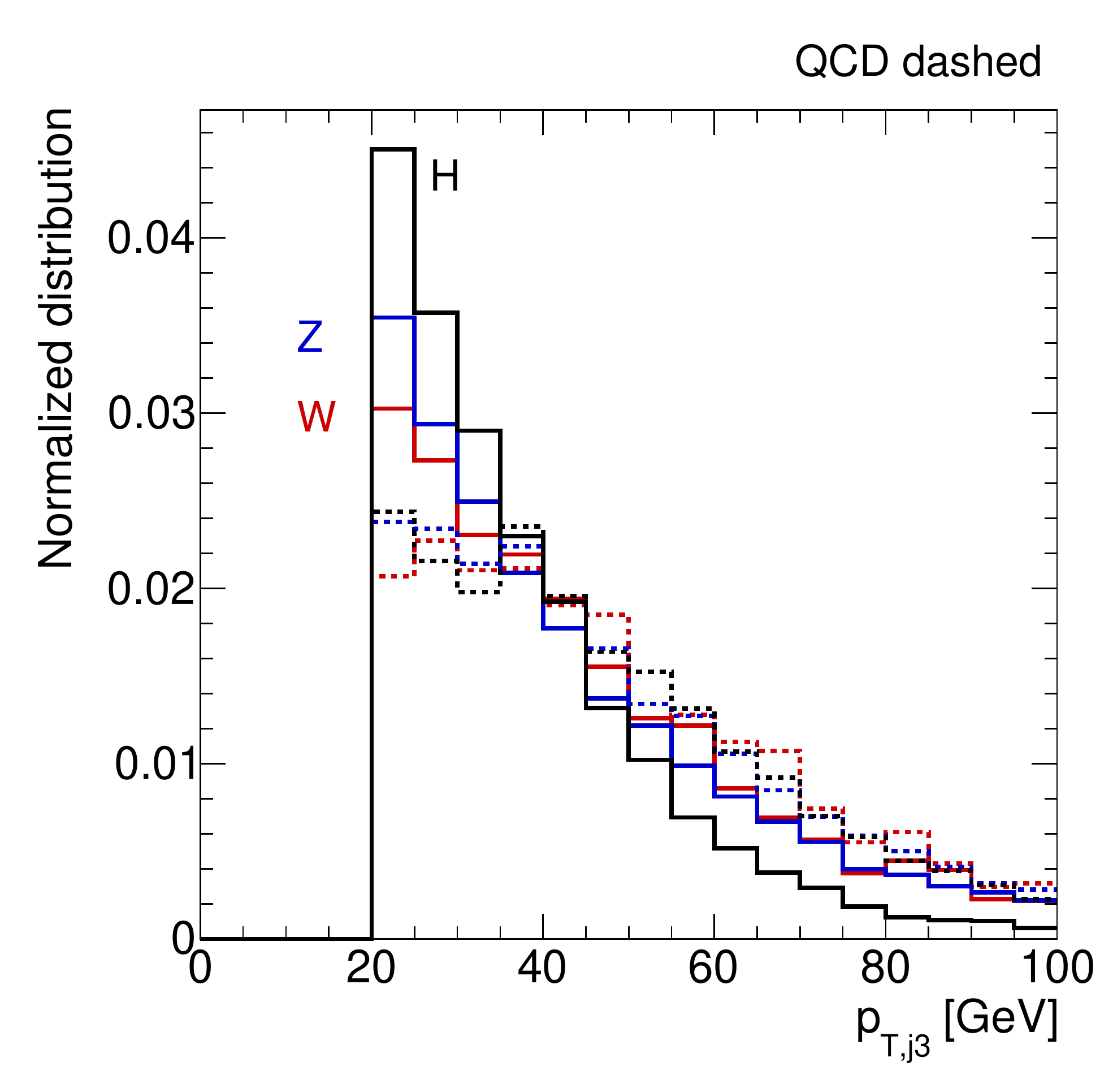}
\caption{Transverse momentum distribution of the three jets for the
  WBF signal and the EW and QCD (dashed) $V$+jets background, after the pre-selection
  cuts of Eq.\eqref{eq:acc_cuts}. }
\label{fig:qg_j3pt_corr}
\end{figure}

Following our two findings, that WBF signal and QCD backgrounds differ
in the partonic nature of the hard jet and that there are PF-level
observables which can separate quarks and gluons, we now apply these
observables to the fully simulated tagging jets.  We employ the usual
\textsc{Sherpa} simulation with two jets for the gluon
fusion signal contribution and up to three hard
jets merged for all other samples 
and a \textsc{Delphes3.3} fast detector
simulation. Unlike for our preliminary results, the combination of
shower and detector simulation no longer defines quark and gluon jets
meaningfully. However, we expect some of the basic parton-level
patterns to remain in the hadron-level final state.  Again using
anti-$k_T$ jets with $R = 0.4$ there is technically no problem in
computing the PF observables defined in Eq.\eqref{eq:qg_obs} after
showering, hadronization and fast detector simulation.

We know that some of the PF distributions are strongly affected not
only by the quark vs gluon nature of the jets, but also by its
transverse momentum. In a realistic simulation of the WBF Higgs signal
and the QCD $Z$+jets background both of these effects are present,
Fig.~\ref{fig:qg_j3pt_corr}. Note that, motivated by the central jet
veto and in the absence of triggering requirements, the third, central
jet can be as soft as $p_{T,j} = 20$~GeV. For the WBF signal where the
tagging jet $p_T$-distribution peaks around 50~...~70~GeV, the
pre-selection cuts of Eq.\eqref{eq:acc_cuts} only have the mild effect
of cutting off the second jet distribution below its peak. The
difference for the QCD background is that typical jet radiation is
neither forward nor at large transverse momentum. The combined tagging
jet cuts of Eq.\eqref{eq:acc_cuts} extract events with harder tagging
jets than we observe for the signal.

In our case the backgrounds are gluon-dominated and harder, so
according to Fig.~\ref{fig:wtrk_dist} the two effects strengthen each
other for $n_\text{PF}$, while they can lead to a compensation for
$w_\text{PF}$. Obviously, a proper multi-variate analysis can
separate the different $p_T$ spectra from the parton nature. On the
other hand, systematic uncertainties might well have a significant
effect on this de-correlation for some of the PF observables.\bigskip

\begin{figure}[t]
\includegraphics[width=0.32\textwidth]{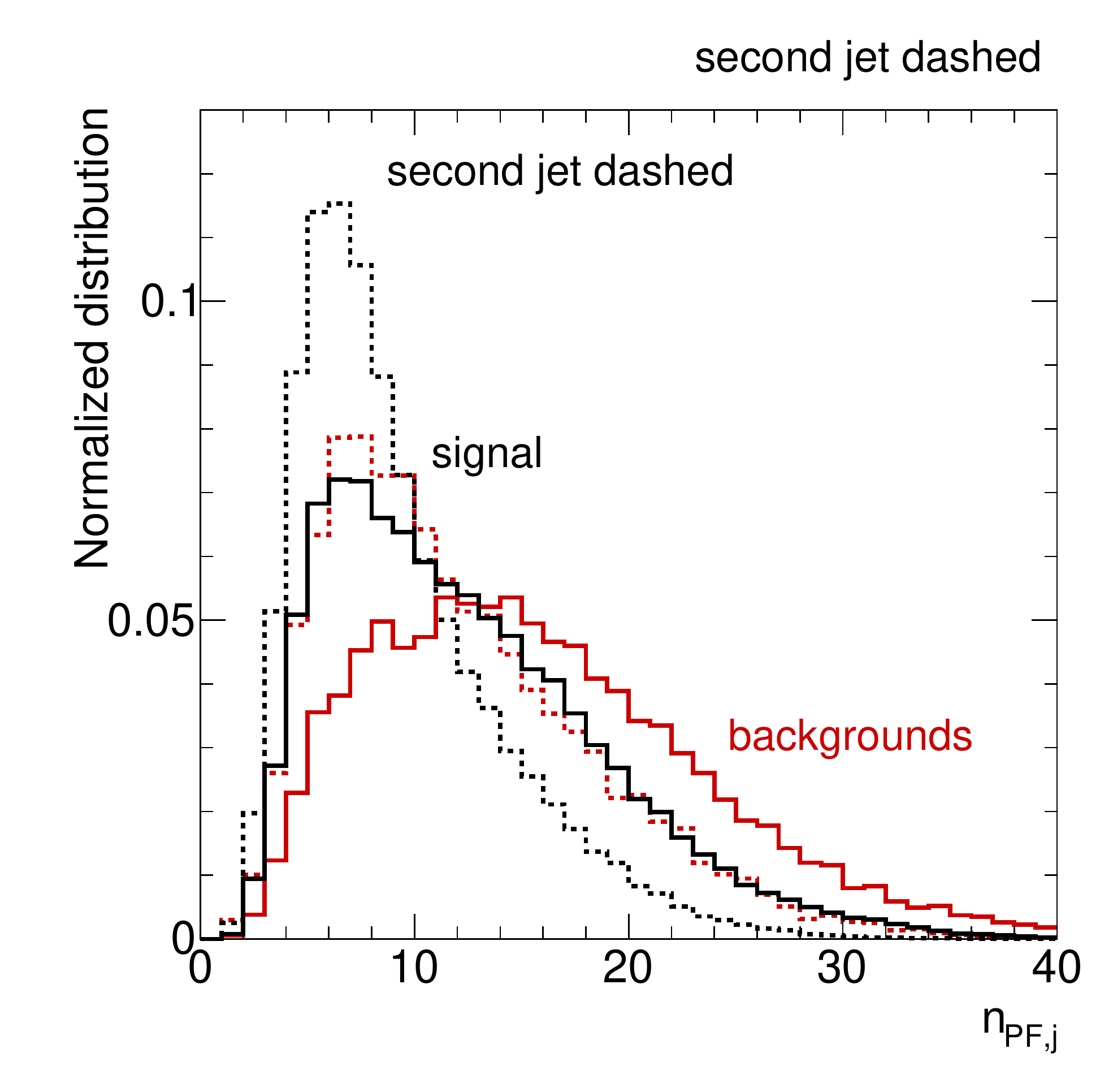}
\includegraphics[width=0.32\textwidth]{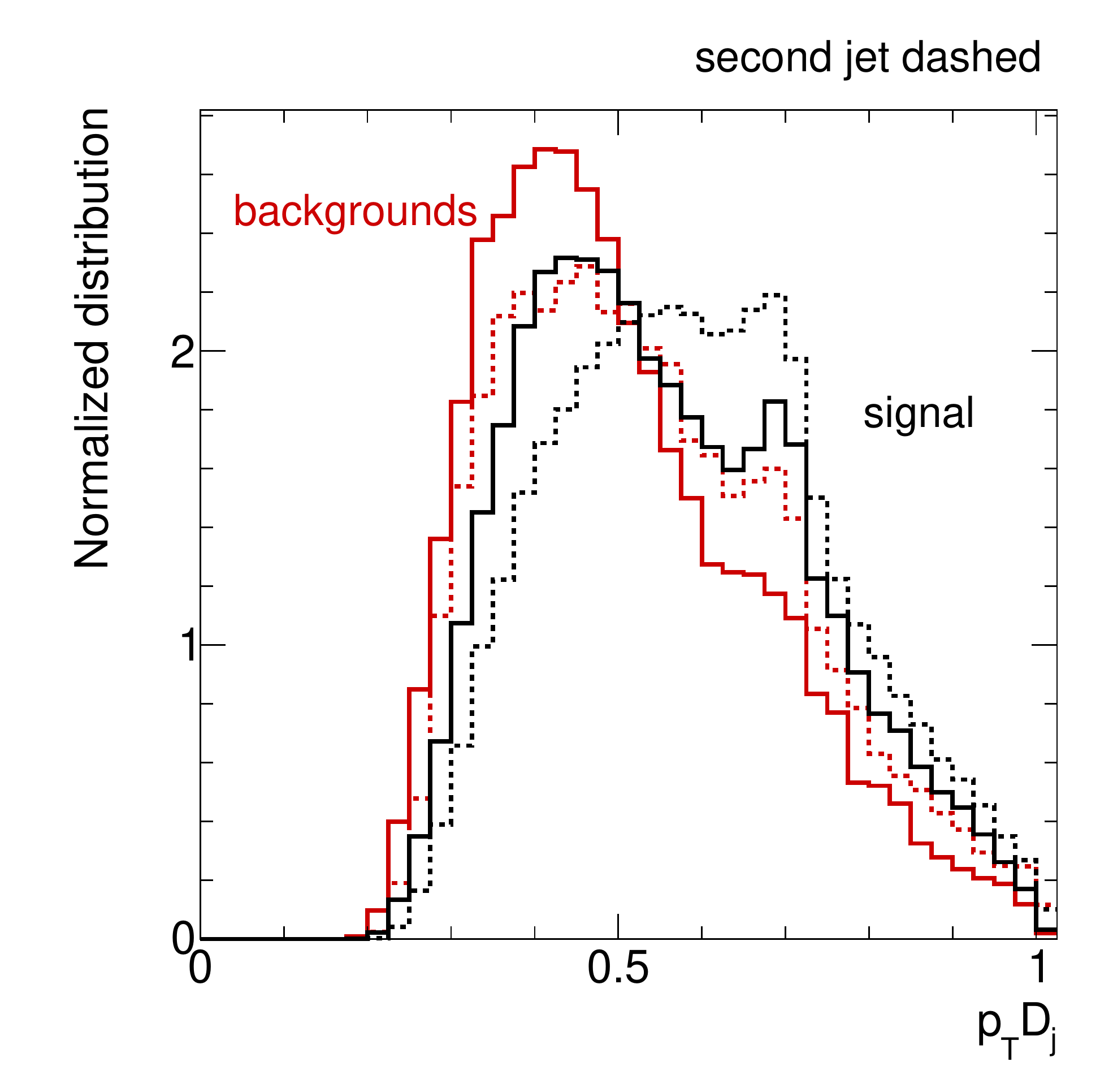}
\includegraphics[width=0.32\textwidth]{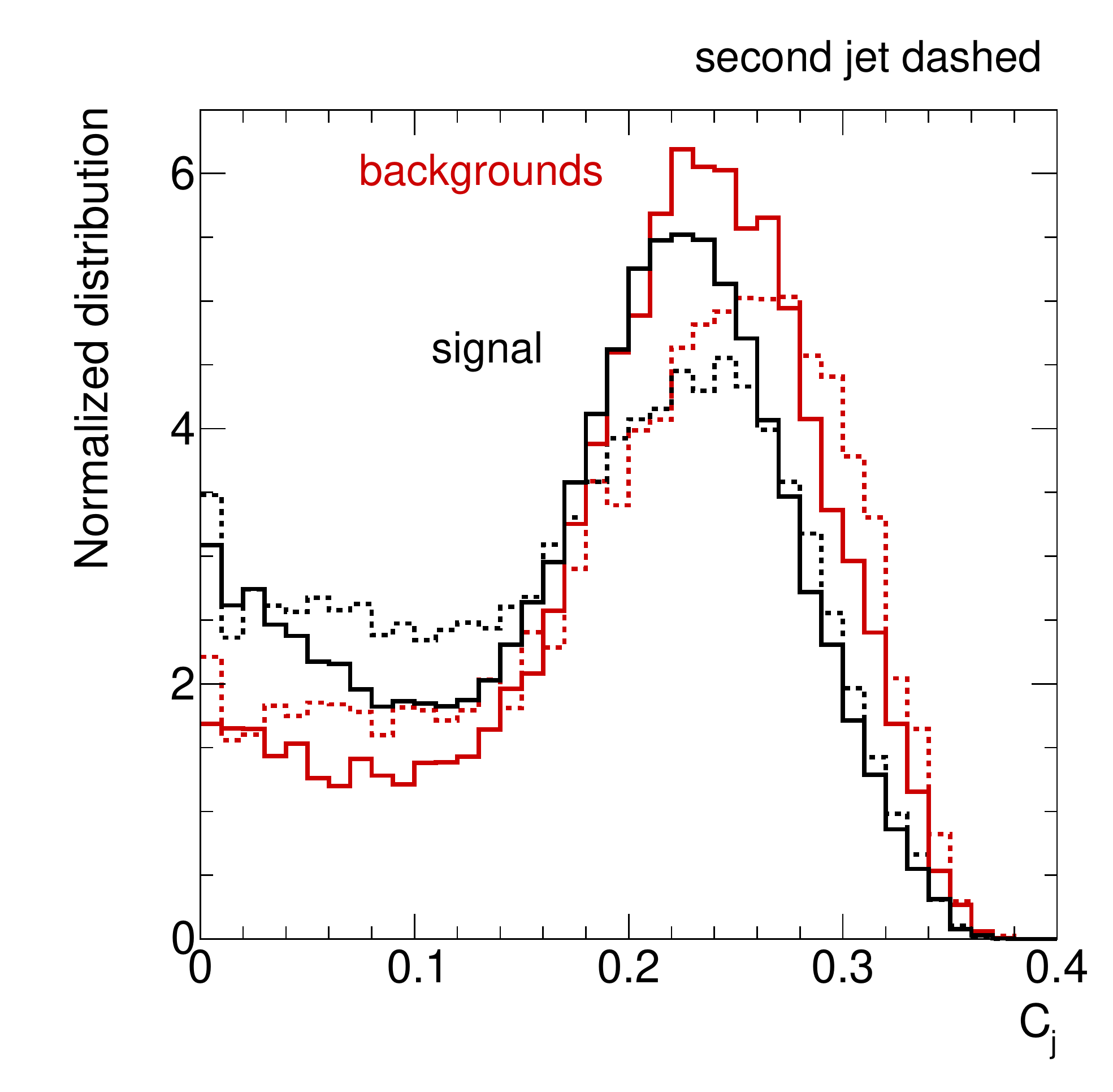} \\
\includegraphics[width=0.32\textwidth]{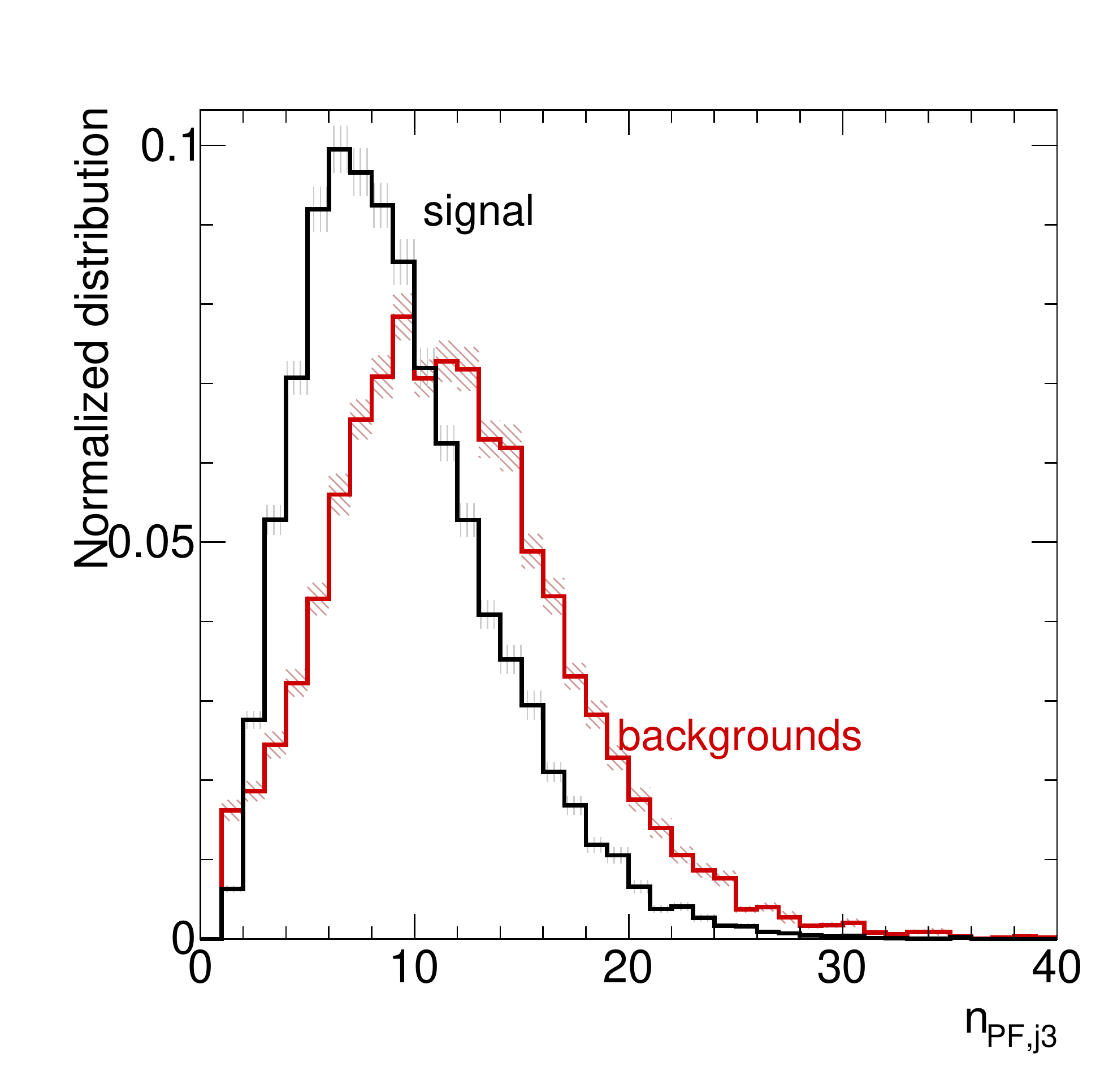}
\includegraphics[width=0.32\textwidth]{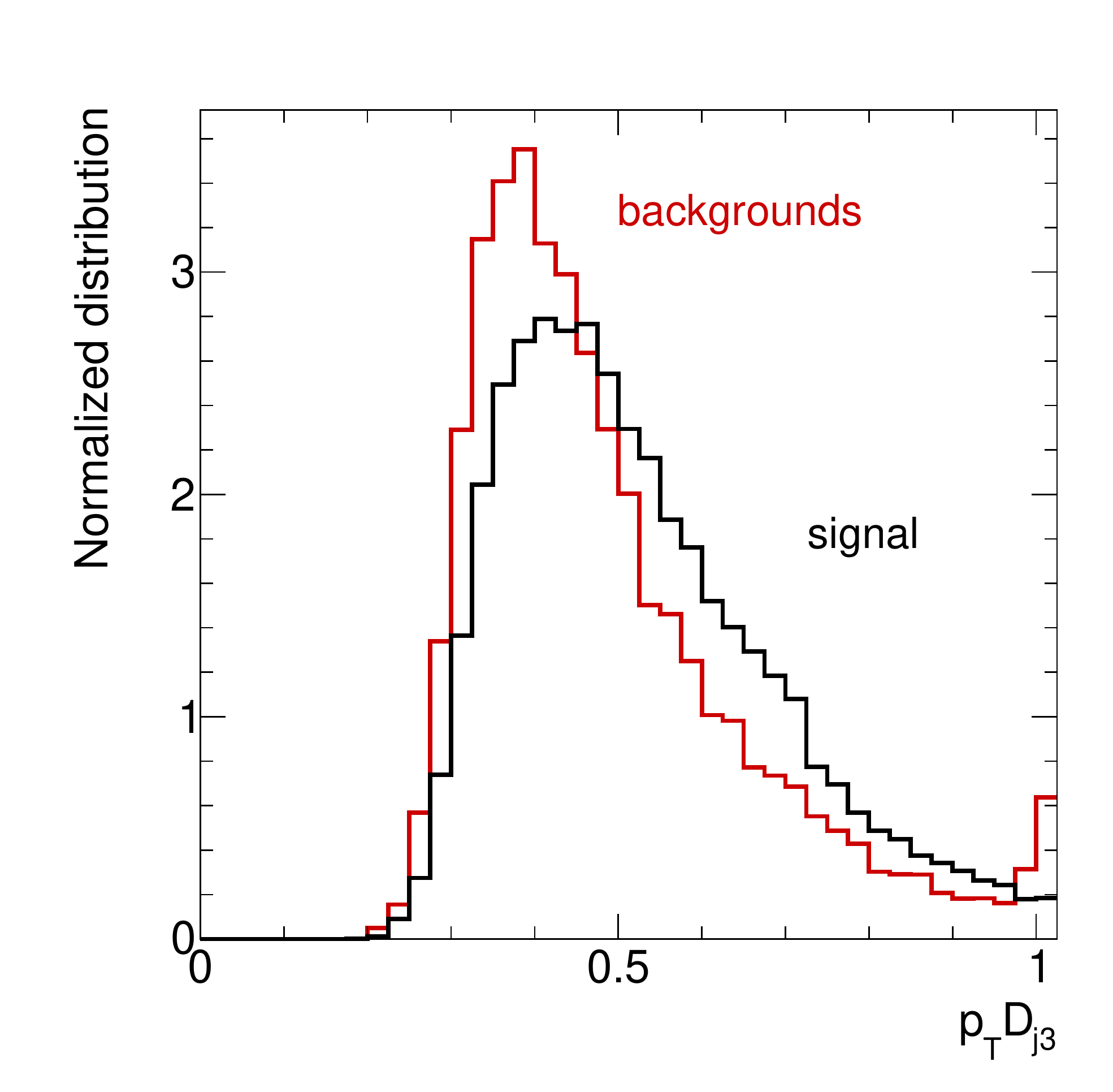}
\includegraphics[width=0.32\textwidth]{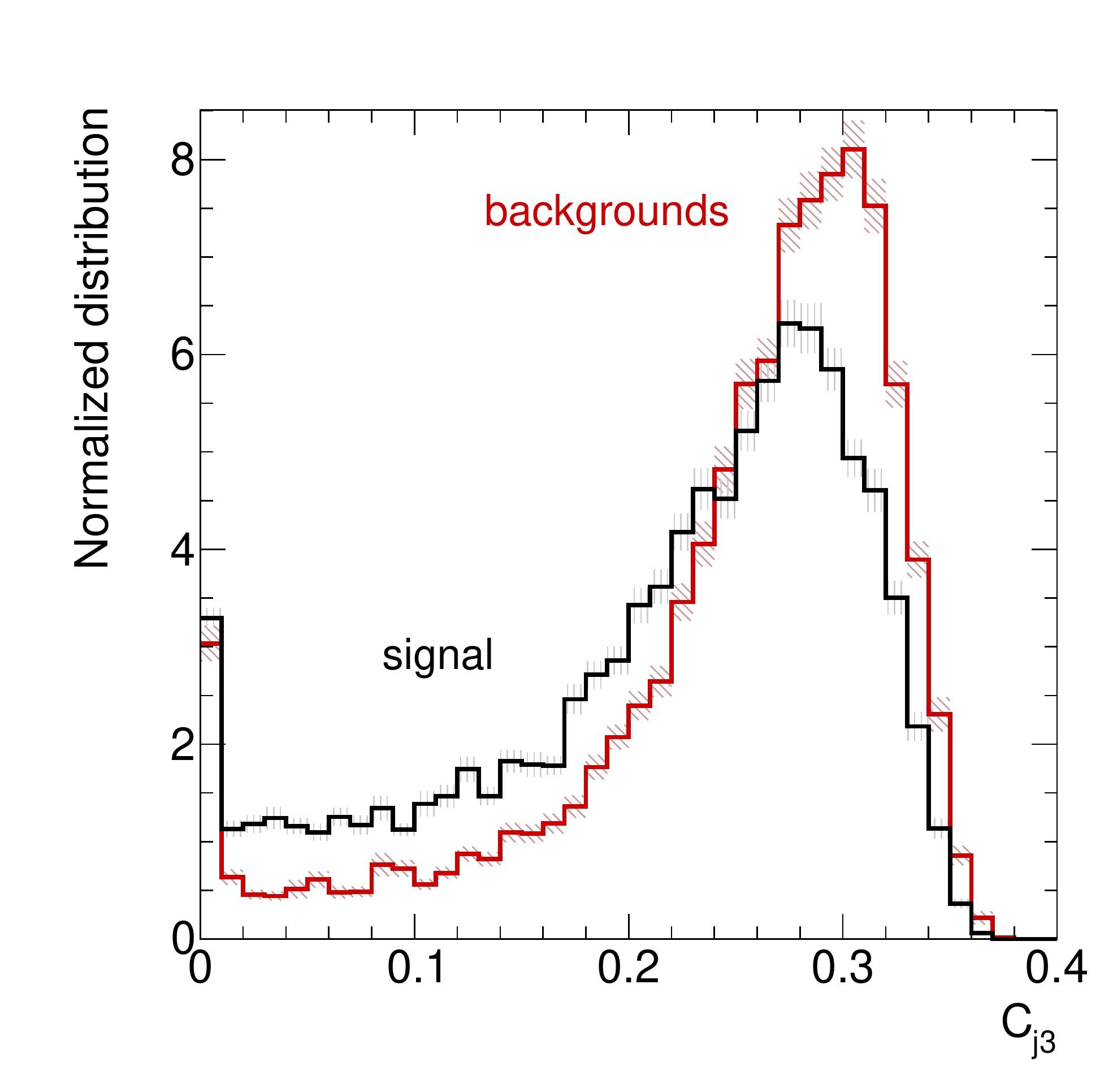}
\caption{Distribution of the quark vs gluon discrimination variables
  $n_\text{PF}$, $p_TD$, and $C$ for the WBF signal and the
  combined QCD and electroweak $V$+jets background, after the
  pre-selection cuts of Eq.\eqref{eq:acc_cuts}. In the top panels we
  show the tagging jets, while in the bottom panels we show the
  softer third jet.}
\label{fig:qg_dist_WBF}
\end{figure}

Finally, in Fig.~\ref{fig:qg_dist_WBF} we display the leading three
observables $n_\text{PF}, p_TD$, and $C$ for the WBF signal
including the contribution from gluon fusion and
the combined $V$+jets background. The two tagging jets are displayed
in the first row, while the second row shows the softer central jet,
which often does not exist for the signal events. The pure number of
PF objects $n_\text{PF}$ shows the best performance for the leading
tagging jet, while $p_TD$ performs better on the second tagging jet.
The reason for this is that the first tagging jet is significantly harder and
hence includes more PF objects.  The second peak in the $p_TD$
distribution for the signal and the electroweak backgrounds is an
artifact of the pre-selection cuts.

For events with three jets we see that the $n_\text{PF}$ distribution
peaks at a value very similar as for the second tagging jet. This is
accidental and an effect of the gluon nature of the jet pushing
$n_\text{PF}$ to larger values and the smaller momentum reducing
$n_\text{PF}$. Still $n_\text{PF}$ as well as $C$ turn out to be
promising observables for the WBF signal extraction.

\section{Performance and triggering}
\label{sec:wbf_trigger}

\def\arraystretch{1.2}
\begin{table}[b!]
\begin{tabular}{lll}
\hline
Set && Variables \\
\hline
jet-level $j_1$, $j_2$ &\quad &    $p_{T,j_1} \quad p_{T,j_2} \quad \Delta\eta_{jj}
	\quad \Delta\phi_{jj} \quad m_{jj} \quad \met
	\quad \Delta\phi_{j_1,\met} \quad \Delta\phi_{j_2,\met}$ \\
subjet-level $j_1$, $j_2$  &\quad &  $n_{\text{PF},j_1} \quad n_{\text{PF},j_2} \quad 
		C_{j_1} \quad C_{j_2} \quad p_T D_{j_1} \quad p_T D_{j_2}$  \\[1mm] \hline
$j_3$ angular information &\quad &   $ \Delta\eta_{j_1,j_3}
	\quad \Delta\eta_{j_2,j_3} \quad \Delta\phi_{j_1,j_3} \quad \Delta\phi_{j_2,j_3}$ \\
jet-level $j_1$-$j_3$ &\quad &   jet-level $j_1$, $j_2$ \quad $+$  \quad 
	$j_3$ angular information \quad $+$ \quad 
	$p_{T,j_3}$ \\
subjet-level $j_1$-$j_3$&\quad &   subjet-level $j_1$, $j_2$ \quad $+$  \quad  
		$n_{\text{PF},j_3} \quad C_{j_3} \quad p_T D_{j_3}$ \\ \hline
\end{tabular}
\caption{Sets of variables used for the BDT analysis.  Variables with
  the subscript $jj$ refer to the two tagging jets.}
\label{tab:bdt}
\end{table}

Using the new handles developed in the last section, we still need to
determine what their impact on the LHC reach for invisible Higgs
decays is.  First, we include the quark-gluon discrimination variables
in a multi-variate analysis. We use boosted decision trees in
\textsc{Tmva}~\cite{tmva,root} with and without treating the quark-gluon
discrimination variables as input parameters for the classification.
We use the AdaBoost algorithm with 70 trees, a maximum depth of 3 and
require a minimum node size of $5 \,\%$ of the number of events.  The
input observables for the BDT analyses are listed in
Tab.~\ref{tab:bdt}.\bigskip

\begin{figure}[t]
\includegraphics[width=0.32\textwidth]{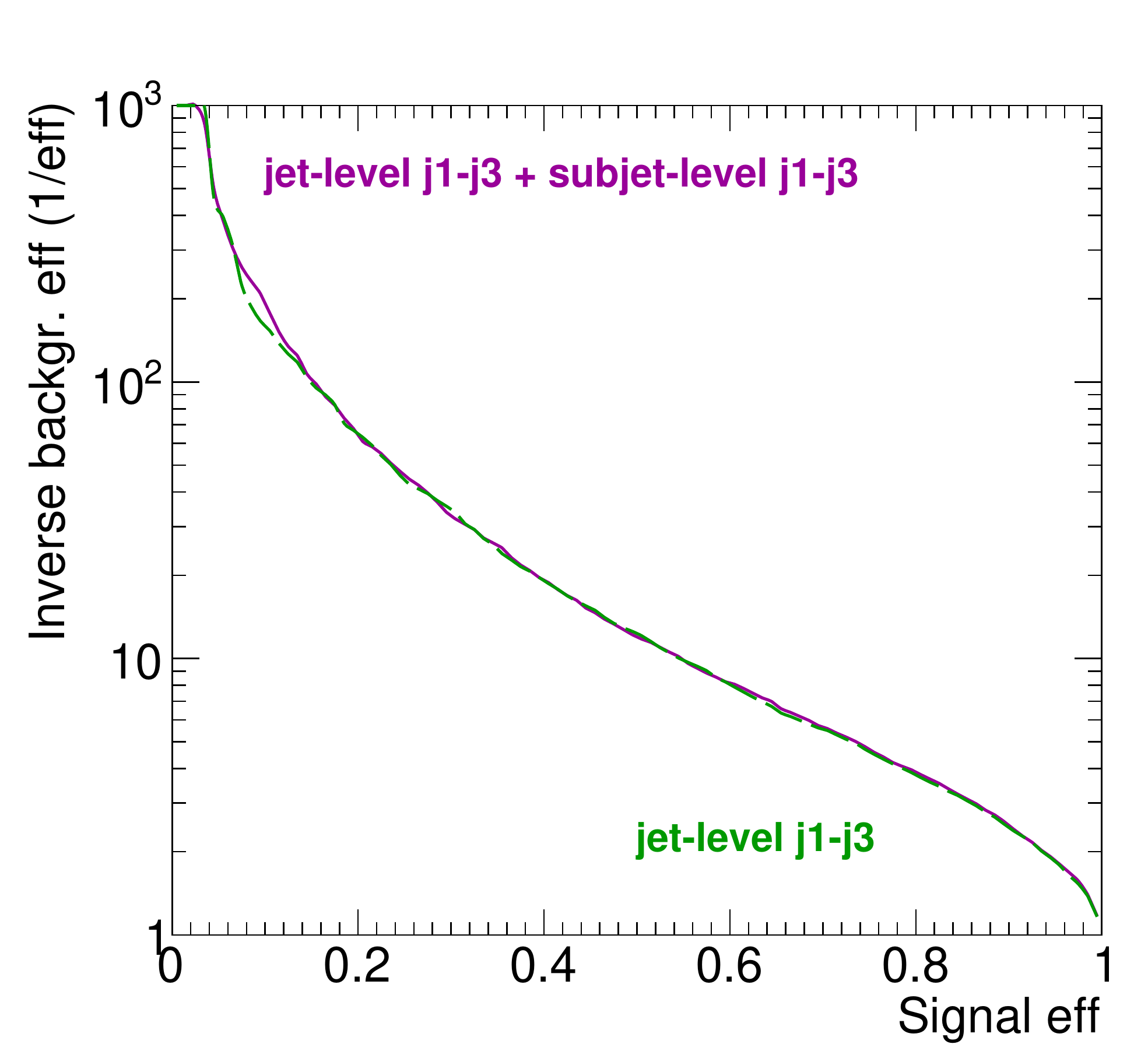}
\includegraphics[width=0.32\textwidth]{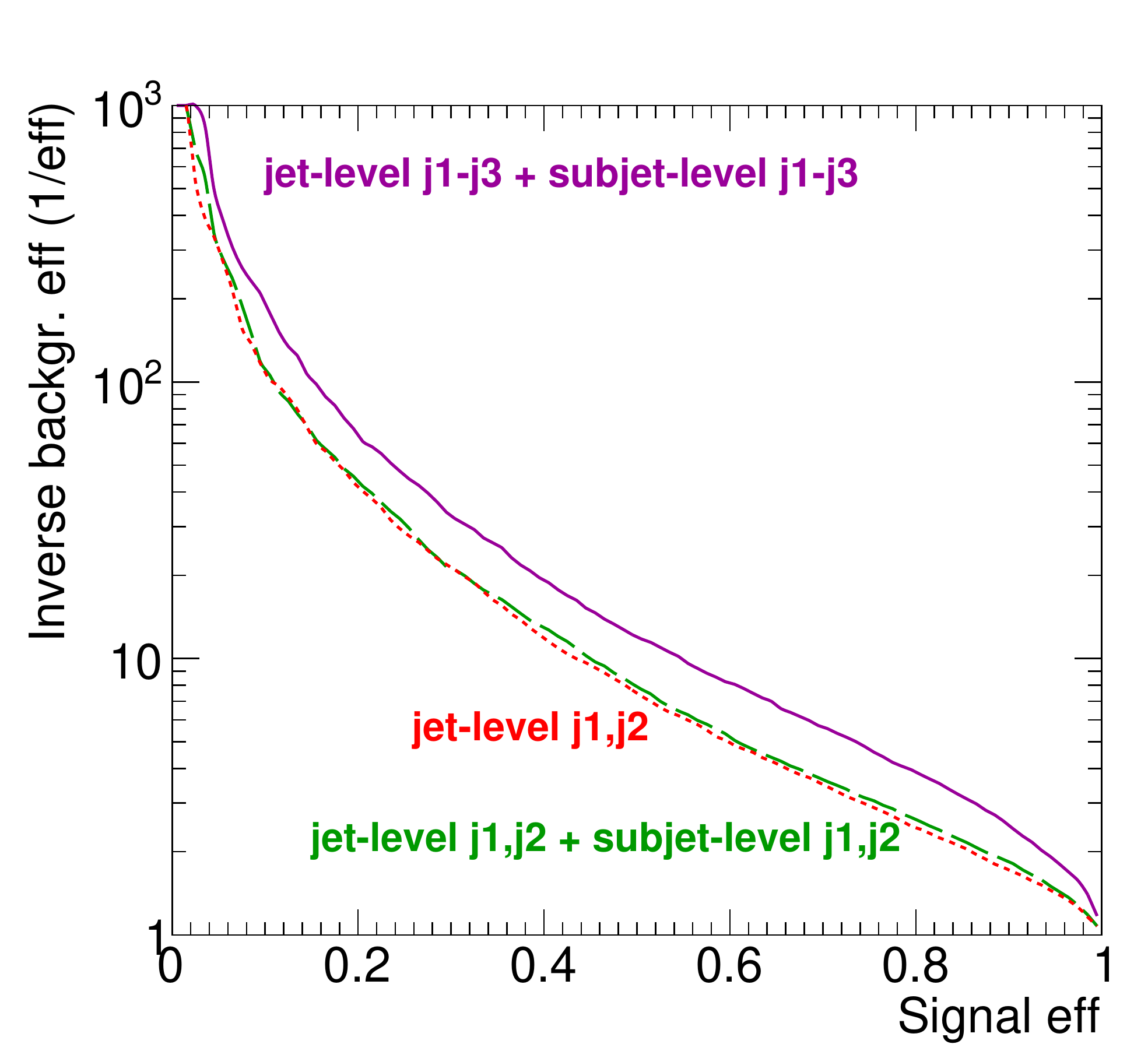}
\includegraphics[width=0.32\textwidth]{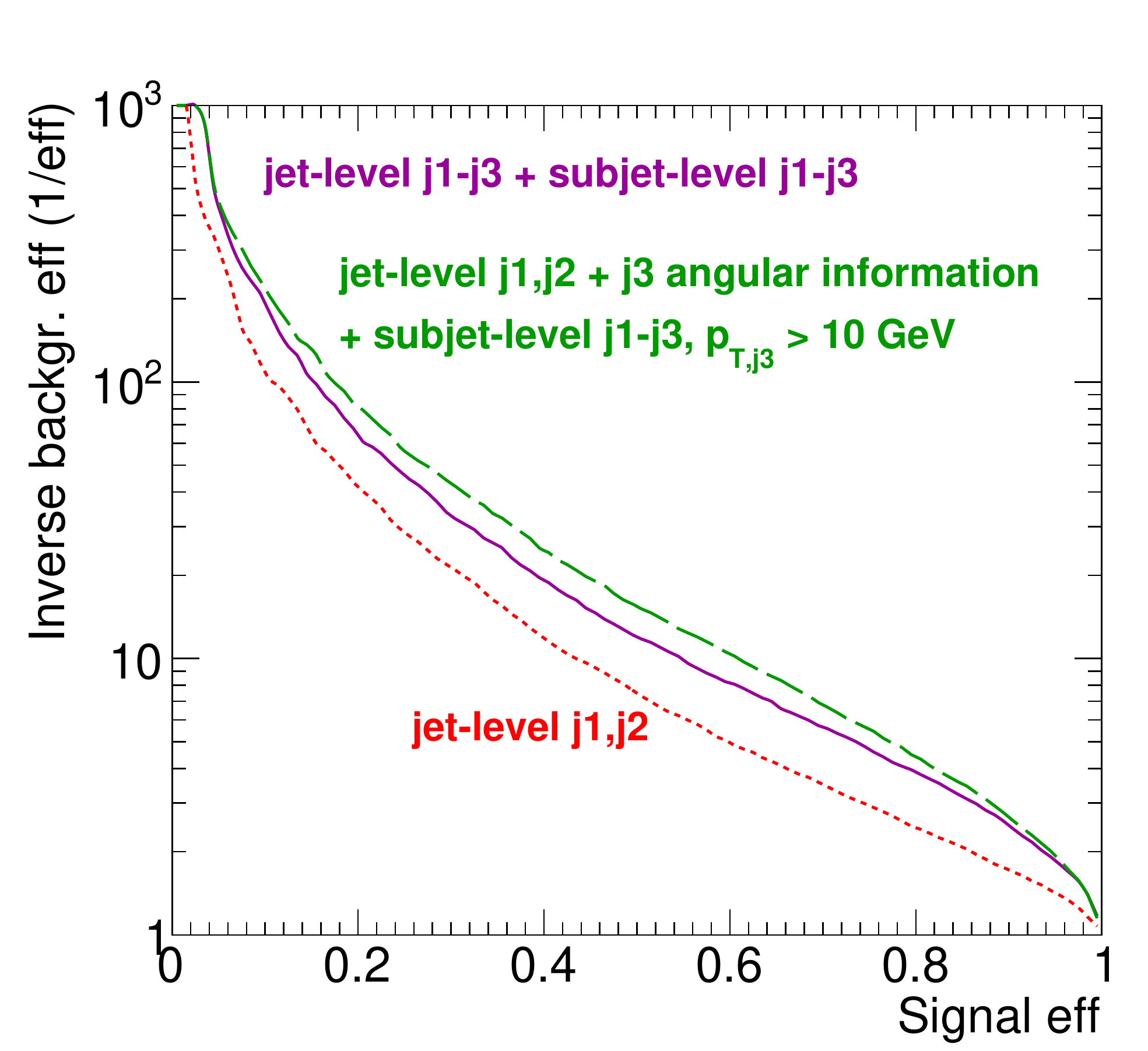}
\caption{Signal efficiency vs inverse background efficiency based on
  jet-level and additional subjet-level information of the leading
  three jets (left). In the center panel we show the jet-level and
  subjet-level of the tagging jets only, while to the right we include
  the subjet-level information on the third jets with a lower
  threshold of $p_{T,j_3}>10$~GeV.}
\label{fig:ROC_WBF}
\end{figure}

In Fig.~\ref{fig:ROC_WBF}, we display the signal efficiency vs inverse
background efficiency (ROC curve) for different sets of BDT input
variables.  The left panel of Fig.~\ref{fig:ROC_WBF} shows the results
from a BDT using the full set of standard WBF variables, containing
the jet-level ($p_T$, $\eta$, $\phi$) information on the tagging jets
and a possible third jet, provided $p_{T,j_3} > 20$~GeV.  We compare
it to a BDT which in addition uses the subjet-level information
$n_\text{PF}$, $C$, $p_TD$. In both cases, the variable most often
used for the splittings of the BDT is $\Delta\eta_{j_1,j_3}$. The most
important subjet-level variables are the number of constituents and
the $p_T D$ of the third jet, individually ranked after the angular
separation variables of the jets.  This ranking corresponds to the
observation that according to the left panel the subjet-level
observables do not lead to a visible improvement of the classification
power.  A similar picture emerges from the center panel, where we show
the classification power based on the two tagging jets alone. Again,
we separate jet-level information alone from the combined jet-level
and subjet-level information.  For the tagging jets the most important
single variable comes out to be $\Delta\phi_{jj}$, while the most
important subjet-level variable is $n_{\text{PF},j_1}$, ranked
fifth. The limited impact of the new set of subjet-level observables
in WBF Higgs production is explained by the large number of
observables shown in Tab.~\ref{tab:bdt}. Even for a $2 \to 3$ process,
the information on the event kinematics is eventually saturated.

From a recent study, we know that the only way to still increase the
performance of the WBF analysis is to use information on softer
central jets~\cite{spying}. The issue with soft central jets is that
they are hard to calibrate. One way around this limitation is to
consider a soft third jet merely a container for subjet-level
observables.  In the right panel of Fig.~\ref{fig:ROC_WBF} we show the
performance once we include jet-level and subjet-level information on
a soft third jet, but omit the jet-level $p_{T,j_3}$.  In this
scenario, the subjet-level discrimination variables can significantly
improve the analysis, with the most important variable becoming
$n_{\text{PF},j_3}$. We have checked that adding the information 
on the jet-level $p_{T,j_3}$ to the input variables
does not further increase the performance of the BDT. 
Including all particle flow observables from
Eq.\eqref{eq:qg_obs} in this analysis is likely overly optimistic. A
proper tracker-focussed analysis with dedicated tunes of simulation
tools is beyond the reach of this first theoretical study, but our
results indicate that such an in-depth analysis is promising.\bigskip

For the HL-LHC, a major issue for WBF Higgs analyses in general and
invisible Higgs searches in particular will be detector limitations
and trigger thresholds. Using the pre-selection cuts of
Eq.\eqref{eq:acc_cuts} is likely overly optimistic.  To systematically
compute the expected limit on invisible Higgs decays for a
multi-variate analysis we use an implementation of the CLs
method~\cite{cls} in \textsc{CheckMate}~\cite{CheckMATE2}.  First, we
scan through different cuts of the BDT classifier variable and
obtain the corresponding signal and background efficiencies. Taking
into account the cross section after our pre-selection cuts of
Eq.\eqref{eq:acc_cuts} we can calculate the $95 \,\%$ CLs limit
for each point on the ROC curve and obtain the best limit.  
We assume a systematic uncertainty
of $3\%$.

\begin{figure}[t]
\includegraphics[width=0.32\textwidth]{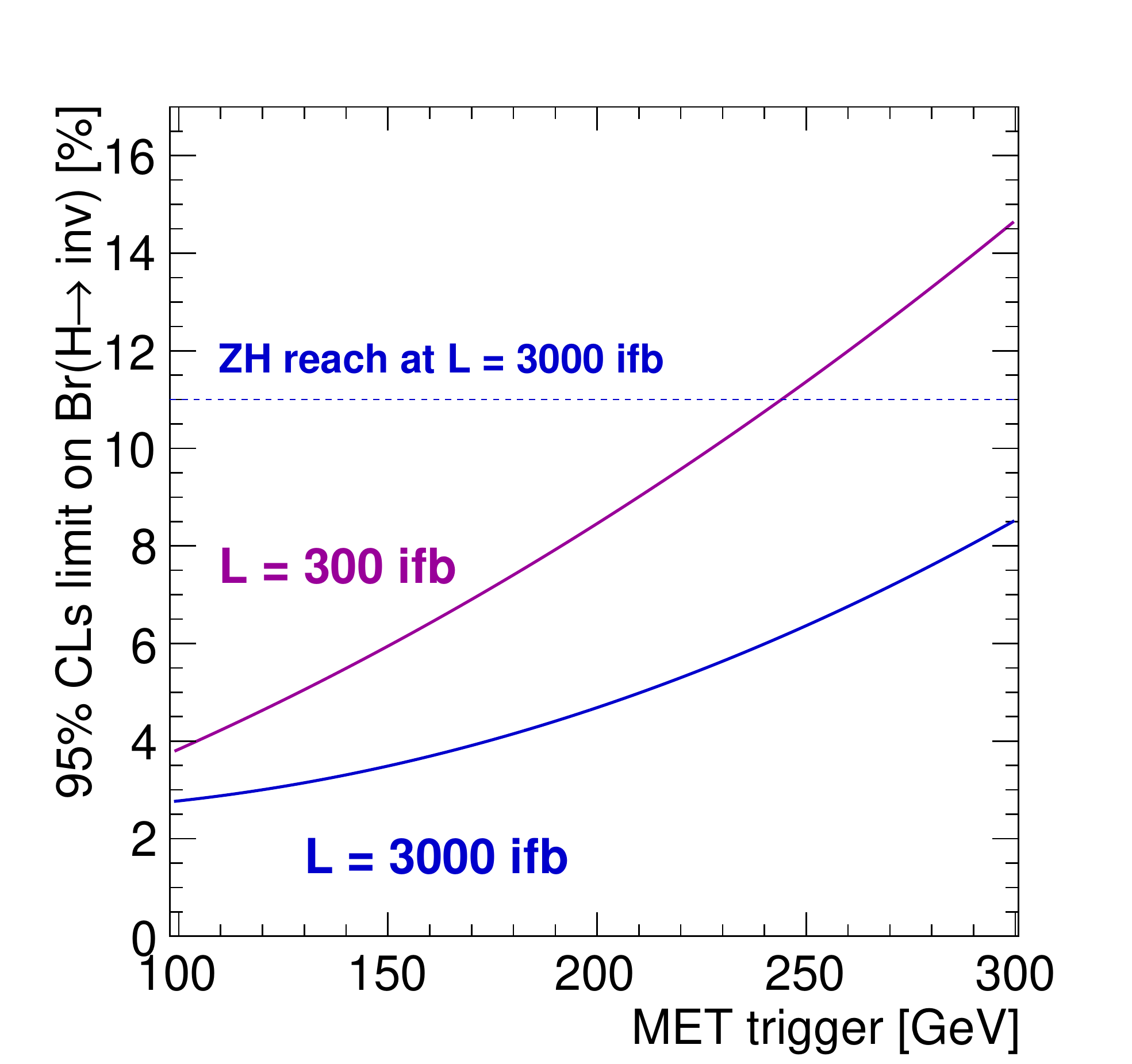}
\includegraphics[width=0.32\textwidth]{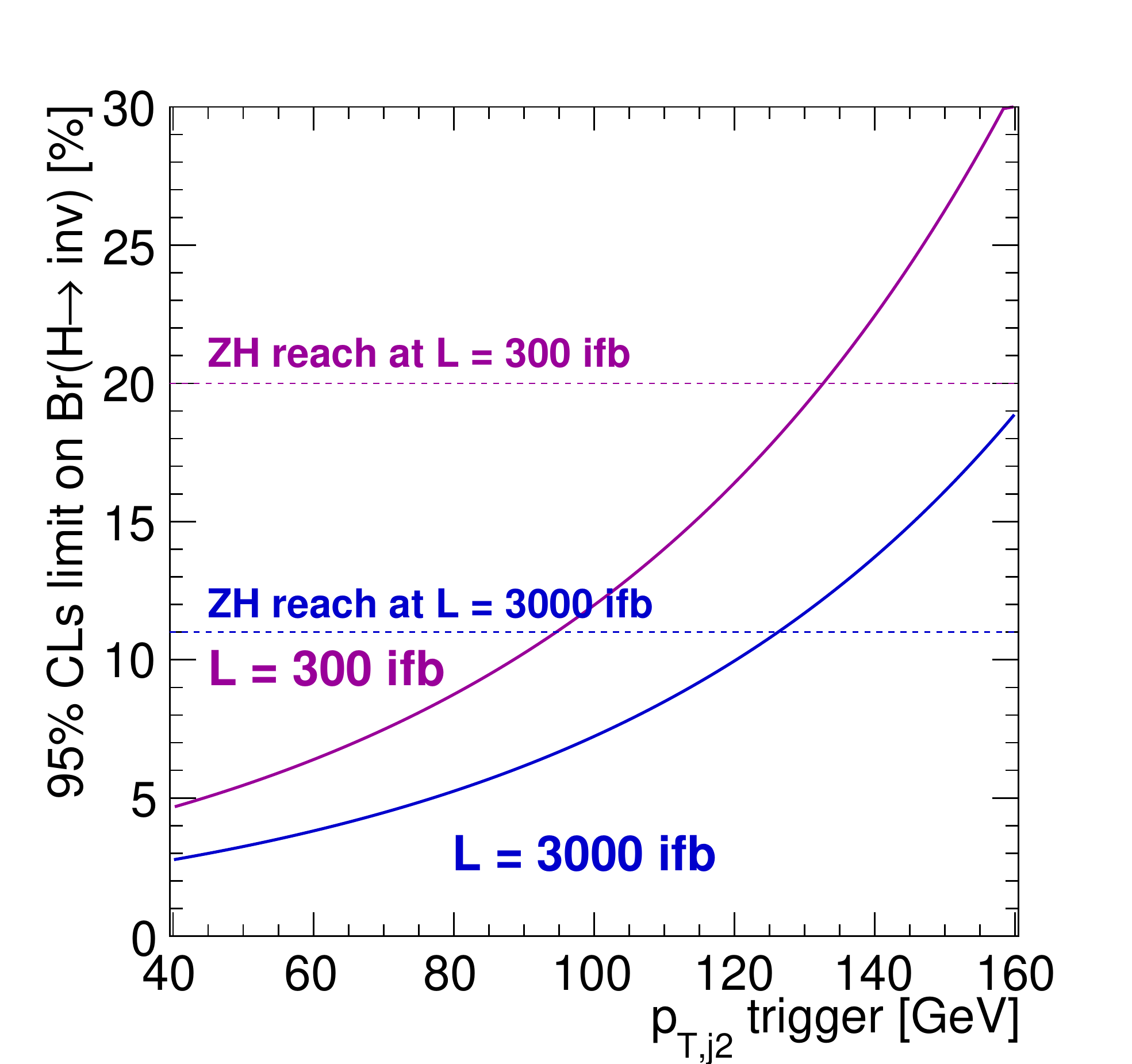}
\includegraphics[width=0.32\textwidth]{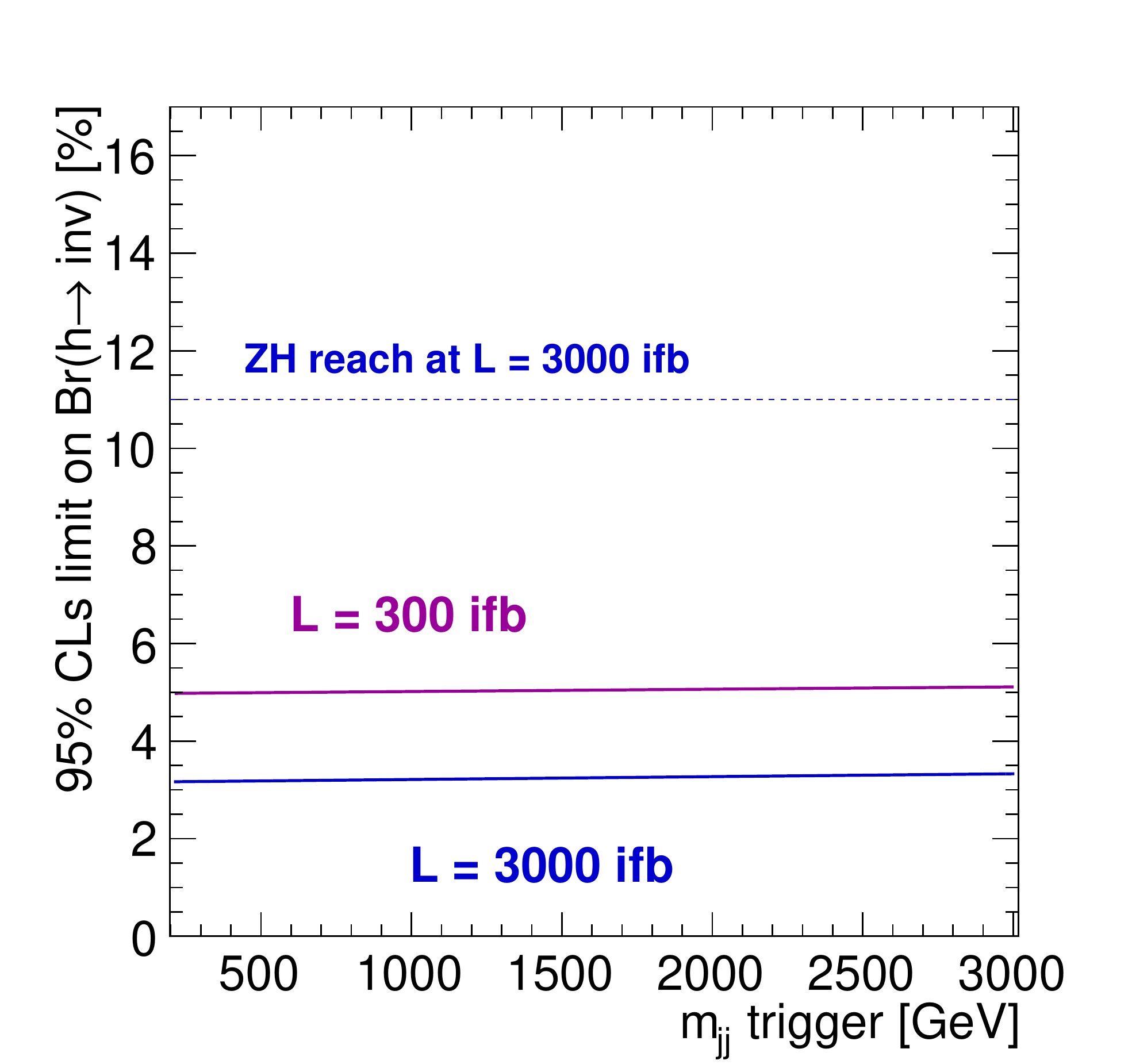}
\caption{CLs limits on invisible Higgs decays from weak boson fusion,
  as a function of trigger cuts on missing transverse energy (left),
  the transverse momentum of the tagging jets (center), and the
  invariant mass of the tagging jets (right). As a reference we also
  display the reach in the leptonic $ZH$ channel, described in the
  Appendix.}
\label{fig:WBF_trigger}
\end{figure}

In
Fig.~\ref{fig:WBF_trigger}, we show the effect of a variation of the
triggers on missing energy, the invariant mass and the transverse
momentum of the tagging jets. As a reference value we also show the
expected results from the leptonic $ZH$ analysis from the Appendix.
While a variation of the minimum missing energy or transverse momentum
has a big impact on the reach for invisible Higgs decays, the
sensitivity seems is largely independent of minimum invariant mass of
the tagging jets, closely related to their rapidity separation.

We find that a trigger cut of $\met > 200~...~300$~GeV
reduces the sensitivity to invisible Higgs decays by roughly a factor
of $1.5~...~2.7$.  While we show trigger-level cuts below $\met >
140$~GeV, we need to keep in mind that in this regime we will likely
have to take into account more background processes.

The trigger cut on the transverse momentum of 
the tagging jets has equally strong impact on the WBF search,
as can be seen already in Fig.~\ref{fig:qg_j3pt_corr}. 
Increasing the requirement to 
around $p_{T,j_{1,2}} > 130$~GeV leads to a drop of the WBF sensitivity to below the $ZH$ reference point.

Finally, the variation of the $m_{jj}$ trigger has almost no impact on
the sensitivity of the analysis. Even for a stiff cut $m_{jj} >
2500$~GeV, the reach in terms of the invisible Higgs branching ratio
stays at the $5\%$ level for an integrated luminosity of $300\,\ifb$
and and at $3\%$ for $3000 \,\ifb$. Since $\Delta\eta_{jj}$ is strongly
correlated with the invariant mass of the tagging jets, we confirm
that an increase in terms of $\Delta\eta_{jj}$ leaves the sensitivity
almost unchanged up to $\Delta\eta_{jj} > 7$.

\section{Outlook}
\label{sec:outlook}

Higgs production in weak boson fusion is the leading process to search
for invisible Higgs decays. The unique tagging jet signature offers
many handles to suppress the different backgrounds, either based on the
actual tagging jets or based on the global QCD structure of the
events.

In this paper we have studied several aspects of WBF tagging jet
analyses. First, we find that backgrounds can indeed be controlled,
but that eventually single top production might limit the
correspondence of $W$+jets and $Z$+jets backgrounds. Next, we have
studied how the dependence of the perturbative signal cross
section prediction on the size of the tagging jets is not specific to
WBF processes and subject to the details of the analysis.

In the main part of the paper we systematically introduced
subjet-level observables for the two tagging jets as well as for
possible central jets. We find that this additional subjet-level
information can be used to suppress backgrounds and increase the
purity of an event sample, but that it does not significantly improve
a multi-variate analysis based on the complete available jet-level
information. The reason is that the kinematics of the signal and
background processes is already fully constrained by the jet-level
analysis. However, we find that subjet observables will allow us to
rely less on notorious observables, like a central jet veto survival
probability. Moreover, subjet information and in particular tracking
information should eventually allow us to exceed the jet-level
performance. Higgs production in weak boson fusion gives us a strong
case to consistently consider jets not as the main analysis objects, but
as containers including valuable subjet-level information.

Finally, we have studied how increased trigger and detector thresholds
will harm the search for invisible Higgs decays at the
HL-LHC. Transverse momentum thresholds can seriously reduce the reach
for invisible branching ratios from the 2\% level to the 10\% level,
\ie the typical reach of the leptonic $ZH$ channel. This is especially
true for the subleading tagging jet. Relying, for example, on the
invariant mass of the tagging jets, in contrast, has hardly any effect
on our WBF analysis.

\section*{Acknowledgements}

The authors would like to thank Monica Dunford and especially Manuel
Geisler for very useful discussions about the discrimination power of
the third jet and for their assistance with \textsc{Tmva}.  We thank
Yacine Haddad for pointing out that the variable $p_{T}D$ is widely
used in the experimental community.  AB acknowledges support by the
IMPRS-PTFS. AB and TP are supported by the DFG Forschergruppe
\emph{New Physics at the LHC} (FOR~2239).

\appendix
\section{ZH benchmark}
\label{app:zh}

\begin{table}[b!]
\centering
\def\arraystretch{1.2}
\begin{tabular}{ccrrr}
\hline
Systematics & \multicolumn{4}{c}{Luminosity $[\ifb]$} \\
            & 36.1~\cite{atlas_zh} & 36.1&300&3000\\
\hline
 1\% sys.&\multirow{2}{*}{39\%} &39\% & 17\% & 8\% \\
 2\% sys.&                       &43\% & 20\% & 11\% \\
\hline
\end{tabular}
\caption{$95 \,\%$ CLs limits on the invisible Higgs branching ratio
  from the leptonic $ZH$ channel. The ATLAS indicated by an asterisk
  result is taken from Ref.~\cite{atlas_zh}.}
\label{zh:trigger:cls}
\end{table}

To make a meaningful statement about the impact of new approaches on
invisible Higgs searches in weak boson fusion, we need a benchmark. We
therefore compute the prospective reach of the associated production
channel
\begin{align}
pp \to ZH_\text{inv} \to \ell^+ \ell^- \; H_\text{inv}
\end{align}
at the HL-LHC. The signature with two same-flavor opposite-sign (SFOS)
leptons plus missing energy is experimentally much more stable than
the WBF tagging jets.  Note that the loop-induced gluon fusion channel
can have sizeable impact on the sensitivity of this
channel~\cite{vh_inv2}. We generate both, tree-level quark-induced and
loop-level gluon-induced events at 14~TeV using \textsc{Sherpa} and \textsc{Delphes3.3}.
We also use \textsc{OpenLoops} for the loop calculations in the 
gluon fusion channels. \bigskip

The main backgrounds are quark-induced and gluon-induced $Z_{\ell
  \ell} Z_{\nu \nu}$ production.  Other important backgrounds are $WZ$
production with a missing lepton from the $W$ decay, $WW$ production
where the invariant mass of the leptons comes out close to the $Z$
mass, and leptonic $t\bar{t}$ production.  We generate these
background processes using \textsc{Sherpa} and include a
loop-level sample for the irreducible gluon-induced $ZZ$ background using
\textsc{OpenLoops}.  All total rates are normalized to their
respective NNLO predictions~\cite{nnlo}.
We require the cuts
\begin{alignat}{9}
p_{T,\ell_1} &> 26 \, \gev & \qqquad 
p_{T,\ell_2} &> 7 \, \gev & \qqquad 
\eta_{e} & < 2.47  & \qqquad
\eta_{\mu} & < 2.5 \notag \\
|m_{\ell\ell} - m_Z | &< 5 \, \gev & \qqquad 
\Delta R_{\ell\ell} &< 1.8 & \qqquad 
\met &> 60\, \gev & \qqquad 
\Delta\phi(p_{T}^{\ell\ell}, \met) &> 2.7 \; .
\label{eq:zh_zll_cuts}
\end{alignat}
Because of the simple $2 \rightarrow 2$ kinematics we do not expect a
large benefit from using a BDT compared to a cut-and-count analysis.
However, to compare the results to our WBF analysis we also analyze
the $ZH$ channels using the BDT implementation of \textsc{TMVA}.  In
addition to the variables given in Eq.~\eqref{eq:zh_zll_cuts} we
include the observables
\begin{alignat}{9}
\left\{ \;
\eta_{\ell_1}, \;
\eta_{\ell_2}, \;
\phi_{\ell_1}, \;
\phi_{\ell_2}, \;
\phi_{\met}, \;
\frac{p_T^{\ell\ell}}{m_T}, \;
N_\text{leptons}, \;
N_\text{jets} \; \right\}     \; .
\label{eq:zh_zll_variables}
\end{alignat}
The resulting $95 \,\%$ CLs limits for a systematic uncertainty of
$1\%$ and $2\%$ are summarized in Table~\ref{zh:trigger:cls}.  While
we assume at best a $2\%$ uncertainty to be realistic, we also show
the results for a $1\%$ uncertainty as a reference.  The comparison
with the expected ATLAS limit~\cite{atlas_zh} at 13~TeV indicates that
appropriate data-driven background rejection techniques can compensate
for otherwise large systematics. The main factor is the normalization
of the leading $ZZ$ background, where we apply a global $K$-factor to
account for the NNLO correction~\cite{nnlo}, while ATLAS uses bin-wise
factors for $m_{ZZ}$~\cite{qqZZ_norm_binwise}.

\newpage
\end{fmffile}

\end{document}

%% file: definitions.tex
\newcommand{\qqquad}{\qquad \qquad}
\newcommand{\qqqquad}{\qquad \qquad \qquad}
\setlist[itemize]{itemsep=3pt,parsep=2pt,topsep=3pt}
\setlist[enumerate]{itemsep=2pt,parsep=2pt,topsep=3pt}

\newcommand{\met}{\slashed{E}_T}

\newcommand{\ie}{i.\,e.\ }
\newcommand{\eg}{e.\,g.\ }

\newcommand{\ifb}{\ensuremath \mathrm{fb}^{-1}}
\newcommand{\iab}{\ensuremath \mathrm{ab}^{-1}}
\newcommand{\gev}{{\ensuremath \mathrm{GeV}}}

\DeclareMathOperator{\br}{Br}

\newcommand{\arxiv}[1]{\href{http://arxiv.org/abs/#1}{arXiv:#1}}